\documentclass[acmsmall]{acmart}

\AtBeginDocument{%
  }

\setcopyright{acmcopyright}
\copyrightyear{2023}
\acmYear{2023}

\acmJournal{TIST}
\acmVolume{Special Issue on Responsible Recommender Systems}

\usepackage{algpseudocode}
\usepackage{algorithm}
\usepackage{caption}
\usepackage{subcaption}
\usepackage{makecell}
\usepackage{tablefootnote}
\usepackage [autostyle, english = american]{csquotes}
\usepackage{wrapfig}
\usepackage[title]{appendix}
\usepackage{xcolor}

\MakeOuterQuote{"}
\newcommand{\MYhref}[3][blue]{\href{#2}{\color{#1}{#3}}}%

\begin{document}

\title[Misinformation Resilient Search Rankings]{Misinformation Resilient Search Rankings with Webgraph-based Interventions}

\author{Peter Carragher}
\affiliation{%
  \institution{Carnegie Mellon University}
  \city{Pittsburgh}
  \state{PA}
  \country{USA}
  \postcode{15217}
}
\email{pcarragh@andrew.cmu.edu}

\author{Evan M. Williams}
\affiliation{%
  \institution{Carnegie Mellon University}
  \city{Pittsburgh}
  \state{PA}
  \country{USA}
  \postcode{15217}
}
\email{emwillia@andrew.cmu.edu}

\author{Kathleen M. Carley}
\affiliation{%
  \institution{Carnegie Mellon University}
  \city{Pittsburgh}
  \state{PA}
  \country{USA}
  \postcode{15217}
}
\email{carley@andrew.cmu.edu}

\renewcommand{\shortauthors}{Carragher, Williams, and Carley.}

\begin{abstract}
The proliferation of unreliable news domains on the internet has had wide-reaching negative impacts on society. We introduce and evaluate interventions aimed at reducing traffic to unreliable news domains from search engines while maintaining traffic to reliable domains\footnote{This article has been submitted to ACM Transactions on Intelligent Systems and Technology for consideration. Its contents may not be further disseminated until a final decision regarding publication has been made and further permission has been granted.}. 
We build these interventions on the principles of fairness (penalize sites for what is in their control), generality (label/fact-check agnostic), targeted (increase the cost of adversarial behavior), and scalability (works at webscale). We refine our methods on small-scale webdata as a testbed and then generalize the interventions to a large-scale webgraph containing 93.9M domains and 1.6B edges. We demonstrate that our methods penalize unreliable domains far more than reliable domains in both settings and we explore multiple avenues to mitigate unintended effects on both the small-scale and large-scale webgraph experiments. These results indicate the potential of our approach to reduce the spread of misinformation and foster a more reliable online information ecosystem. This research contributes to the development of targeted strategies to enhance the trustworthiness and quality of search engine results, ultimately benefiting users and the broader digital community.

\end{abstract}



\maketitle

\section{Introduction}

The impact that search engines have on exposure to misinformation is still poorly understood, but the effect has been felt intimately by many. In 2015, Dylan Roof, a neo-nazi and white supremacist, opened fire in a Bible Study taking place at Emanuel African Methodist Episcopal Church. He killed 9 people---all African American. In the manifesto he posted online, he attributed his "awakening" to the first time he typed the words "Black on White crime" into Google. He recounts that the first website he discovered while searching was "The Council of Conservative Citizens"\footnote{\url{conservative-headlines.com} appears as a link-scheme site whose impact on the search rankings we aim to reduce.}---a misinformation news site\footnote{\url{https://mediabiasfactcheck.com/council-of-conservative-citizens/}} operated by a white supremacist hate group\footnote{\url{https://www.adl.org/resources/news/extremism-america-council-conservative-citizens/}}. Roof writes, "I have never been the same since that day" \cite{dylanroofgoogle}. Combating and better understanding the proliferation of misinformation on search engines is an area of fundamental importance.

Measuring the causal impact of unreliable news exposure is challenging given the complexity of information consumption ecosystems, but numerous studies have highlighted ways that unreliable news can erode public trust in media and science. Exposure to unreliable news has been linked to lower trust in media, with higher exposure to unreliable news being associated with higher distrust in media \cite{ognyanova2020misinformation,wasserman2019exploratory}. Exposure to COVID-19 misinformation online has been strongly associated with lower COVID-19 vaccination rates \cite{loomba2021measuring,neely2022vaccine}, and exposure to even a small amount of climate change misinformation can lower acceptance of climate change \cite{cook2022understanding}. The vast majority of online misinformation and unreliable news research focuses on the spread of misinformation through social media platforms, but search engines like Google, Bing, Yahoo, DuckDuckGo, and many others, also play an important, albeit understudied, role.

Research into how search engines spread misinformation has been primarily based on "audit" studies, where researchers query a set of pre-selected keywords, sometimes on multiple search engines, and label the reliability of the top results \cite{brady2021social}. Audit studies can provide answers to challenging questions. For example, Makhortykh et al. used search engine audits to assess whether Google and Yandex were complying with a court order to censor certain queries in Russia \cite{makhortykh2022story}, and Urman et al. used search engine audits to analyze the degree to which various search engines return results that promote conspiracy theories \cite{urman2022earth}. However, audit studies are constrained by the set of keyphrases chosen by the authors of the study, so it is unclear how well they generalize to the true dynamics of unreliable news exposure. Golebiewski and Boyd highlight the impact of "Data Voids"---keyphrases for which relevant returns are sparse---in the broader misinformation ecosystem \cite{golebiewski2019data}. As search engines assume that each query has a relevant response, highly-specific keyphrases built around conspiracy theories have very little competition from reliable domains\footnote{For example, as of June 2023 the keyphrase "pope red shoes human skin" ranks highly for two unreliable domains, stillnessinthestorm.com and prepareforchange.net}. Recent research has found strong evidence that certain types of black-hat SEO are particularly common in known misinformation domains \cite{williams2023search, misinformation_detection}. To mitigate the impacts of link farming on driving unreliable traffic, we propose four principles for webgraph ranking interventions: fairness, generality, cost, and scalability.

In line with these principles, we design and test various interventions on small-scale webgraph data extracted from SEO toolkits, and highlight this data's usefulness as a developmental testbed. On the small-scale data, we use reliability-labeled news site traffic estimates to design and assess the effectiveness of two intervention classes aimed at reducing traffic to unreliable domains while maintaining traffic to reliable domains. We evaluate the small-scale webgraph interventions by using a regression model to predict changes in traffic based on simulated modifications in the backlink networks of labeled news domains. We then generalize the most effective class of interventions and test the effectiveness of these interventions on a large-scale webgraph consisting of 93.9 million domains with 1.6 billion edges by calculating the relative change in each domain's PageRank. We evaluate our intervention against variants of the well-known Personalized PageRank algorithm and find that incorporating Anti-TrustRank scores improves results. We open-source all code\footnote{\url{https://github.com/CASOS-IDeaS-CMU/Misinformation-Resilient-Search-Rankings}} and data\footnote{\url{https://doi.org/10.1184/R1/25174193.v1}} used in this paper. To our knowledge, this is the first work that demonstrates the usefulness of small-scale webgraphs in designing webspam interventions that are then tested on large-scale webgraph data. 

In the small-scale webgraph experiments, our best intervention yields a substantial 35\% and 27\% drop in the predicted traffic and ranking of unreliable domains respectively. In contrast, this intervention yields a much lower 11\% and 8\% drop in predicted traffic and ranking for reliable domains. These results demonstrate a favorable trade-off between reducing the influence of misinformation domains and minimizing collateral damage to reliable domains. In the large-scale experiments, we find further evidence that the proposed interventions lead to a favorable tradeoff, albeit with reduced effects---removing link scheme domains leads to a drop in PageRank centrality of 10\% for unreliable domains and 2\% for reliable domains. Finally, we explore the unintended effects of our interventions on both small-scale and large-scale webgraphs and propose several mitigations. In doing so, we develop a more refined definition of link scheme services as domains that serve to inauthentically increase traffic to target sites by embedding hundreds of thousands of low-quality links into their HTML with little context. As a result, link scheme providers may distribute their links to target domains of various unrelated categories, creating a unique linking pattern that we use to increase the precision of link scheme identification algorithms, at the cost of recall.

To our knowledge, this is the first work that attempts to penalize unreliable news domains within search engines. Combating attempts of fake news sites to manipulate search engines presents unique opportunities for interdisciplinary collaboration, as the goals of search engine companies, ad tech companies, regulators, and other stakeholders are aligned in this space. Search engine companies have monetary interests in stopping search manipulation, ad tech companies have a monetary interest in not serving ads to companies manipulating traffic, and many stakeholders have an interest in providing reliable information to users and the electorate \cite{ghosh2018digital}. We hope that this encourages future work to reduce the dissemination of misinformation via search engines. 





\section{Related Work}
\subsubsection{PageRank}


With the advent of the World Wide Web in the 1990s, Kleinberg \cite{kleinberg_pagerank} developed an iterative approach for determining authoritative sources on a given topic. In line with this work, Page et. al. \cite{google_pagerank} proposed the PageRank algorithm which served as the basis for information retrieval in early search engines. PageRank works by scoring websites using link quantity and link quality. Search engines now use many metrics to rank results, which are not publicly available, but PageRank and its core intuitions are still widely considered to be important metrics. Reports on a 2023 Yandex source code leak found that the original PageRank and related metrics were still factored into rankings \cite{yandexpagerank}. 

Since the advent of Google's PageRank, an entire industry centered around gaming its metrics has emerged. This can be seen as an example of Goodhart's law \citep{goodhart}; "any observed statistical regularity will tend to collapse once pressure is placed upon it for control purposes." Put simply, "when a measure becomes a target, it ceases to be a good measure." In this case, Search Engine Optimization (SEO) has become a canonical example of Goodhart's `control pressures'.

\subsubsection{Search Engine Optimization}
The SEO industry is valued at 80 billion dollars and focuses on increasing the traffic that websites receive through search engines \citep{seo80bil}. SEO usage has become ubiquitous, with a study showing that 58\% of over 250,000 search results were certainly optimized \citep{seo_prevalence}. Although search engines support numerous SEO practices, it's important to note that many link scheming and spamming practices \cite{black_hat_seo_review} intended to inauthentically manipulate recommendation algorithms are prohibited by search engine webmaster guidelines \footnote{https://developers.google.com/search/docs/essentials/spam-policies}. 


\subsubsection{Link Spam Detection}
Link spamming (or link scheming) is likely the most common "black-hat" SEO tactic \citep{black_hat_seo_review}, and many authors have proposed algorithms and models to address the problem of link spam detection. Despite the continued importance of identifying and penalizing webspam influence, relatively little work has been done in this space since the early 2000s. This line of work began with TrustRank \cite{gyongyi2004combating}, a clever extension of Personalized PageRank (PPR) \citep{jeh2003scaling}. The family of algorithms and extensions built around TrustRank remains highly influential. 

\textbf{Personalized PageRank} (PPR) tailors rankings to specific set of domains. These domains form a preference vector that is used to initialize PageRank (PR) scores for each site before running the PR algorithm. By incorporating the preference vector, PPR computes rankings that reflect both the inherent importance of pages and their relevance to the preferred domains \citep{jeh2003scaling}. 

One of the first, and arguably the most well-known algorithms for combating web spam is TrustRank \cite{gyongyi2004combating}. TrustRank seeds the PPR preference vector with a set of reliable domains. In turn, PPR spreads trust from those domains out to the broader network. An effective modification was later proposed called \textbf{Anti-TrustRank} (ATR) that uses a similar set-up, but starts with a set of unreliable domains and reverses the direction of all the edges in the webgraph before running the PPR algorithm \cite{jiyoung_whang_scalable_2020}. The authors demonstrate that this approach is effective in finding link spam sites, which tend to heavily link to one another. The ATR algorithm has seen continued development \citep{whang_fast_2018, jiyoung_whang_scalable_2020} and remains one of the most well-known domain-level spam-detection algorithms. 


\subsubsection{Unreliable News and Link Spam} 


Only a handful of studies have explored the impact of search engine rankings on the news ecosystem, with the majority of misinformation research focusing on social media \cite{sharma2019combating}. Williams and Carley \cite{williams2023search} showcased the use of webgraph data to identify large-scale link scheme activities associated with Kremlin-aligned propaganda domains. Similarly, Bradshaw \cite{bradshaw2019disinformation} conducted an analysis of how 29 junk news sites optimized keywords over three years to increase web traffic and spread disinformation. Urman et. al. \cite{urman2022earth} explore conspiratorial searches conducted on Google, Bing, DuckDuckGo, and Yandex, revealing consistent misinformation in the top search results of Yandex and DuckDuckGo. 

Several prior studies explored the predictive power of webgraphs on tasks related to misinformation domain detection \citep{www_bias_detection, hrckova2021quantitative, link_scheme_misinfo}. Notably, Carragher, Williams, and Carley investigate the role that spam-based black-hat link-building techniques have on spreading misinformation \citep{misinformation_detection}. This work develops an algorithm for identifying link scheme sites among backlinking domains as those that disproportionately link to unreliable news domains (\autoref{fig:link_scheme_target_dist}). We use this approach in our proposed link scheme removal intervention.

\subsubsection{Web Traffic Estimates}
Traffic is the clearest signal through which search engines impact websites. Unfortunately, unless site owners elect to publicly share this information, only site owners know the actual levels of traffic their sites get. However, given the importance of SEO strategies in site performance, commercial data sources for traffic estimates are readily available from large-scale SEO toolkit companies. 

A non-peer-reviewed blog post by AuthorityHacker compared traffic estimates from various SEO toolkits with actual traffic estimates provided by 47 website owners and found Ahrefs to be the most accurate of any SEO toolkit \cite{ahrefstraffic}. Ahrefs self-reported similar results across a larger set of 1,635 domains \footnote{https://ahrefs.com/blog/traffic-estimations-accuracy/}. Ahrefs bases its traffic estimates on position-weighted click-through rate estimates for Google search volumes of all keyphrases for which a website appears in the top 100 Google search results. While we do not expect these traffic estimates to be completely accurate, Ahrefs had the 5th most active commercial webcrawler in June 2023 by number of requests according to Cloudflare Radar\footnote{https://radar.cloudflare.com/traffic/verified-bots}, lending their traffic estimates further credence. As a further quality control check, we compare Ahrefs traffic estimates with SimilarWeb \citep{similarweb} traffic estimates and CommonCrawl PageRank \citep{commoncrawl} in Section \ref{sec:WTEData}.

\subsubsection{Fair PageRank}
A wealth of research has been done on improving the fairness of ranking algorithms in terms of protected attributes \citep{pitoura_fairness_2022, tsioutsiouliklis_fairness-aware_2021}. While this is a crucial line of study, we do not find extensive works in the literature attempting to design interventions that promote search engine safety and resilience to misinformation. Furthermore, principles of fairness and safety may run counter to one another. Given that censorship and misinformation interventions have become polarizing topics, we believe it is important for researchers to encourage public trust and acceptance by designing search engine ranking systems that are both safe and fair.

\section{Data}

\newcommand\urlpairs{446,161}

\subsection{Misinformation Labels} 
We utilize a pre-existing dataset of 3211 news domains with accompanying reliability and political bias labels \footnote{\url{https://kilthub.cmu.edu/articles/dataset/Dataset_for_Detection_and_Discovery_of_Misinformation_Sources_using_Attributed_Webgraphs_/25174193/1}} \citep{misinformation_detection}. This list is composed of other previously published domain lists \citep{blacklist, original_list}, is primarily based on mediabiasfactcheck.com (MBFC)\footnote{\url{https://mediabiasfactcheck.com/methodology/}} \citep{mbfc}, and has substantial overlap with recent works investigating consistency of news reliability ratings across annotators \citep{domain_rating_correlation}. Our list does not include proprietary NewsGuard\footnote{\url{https://www.newsguardtech.com/}} ratings as there is less data to compare their annotations to---only about 20\% of the 8,178 publicized NewsGuard domain ratings have been rated by MBFC \citep{domain_rating_correlation}.



\begin{figure}
\centering
\begin{subfigure}{.45\textwidth}
    \centering
    \includegraphics[width=.95\linewidth]{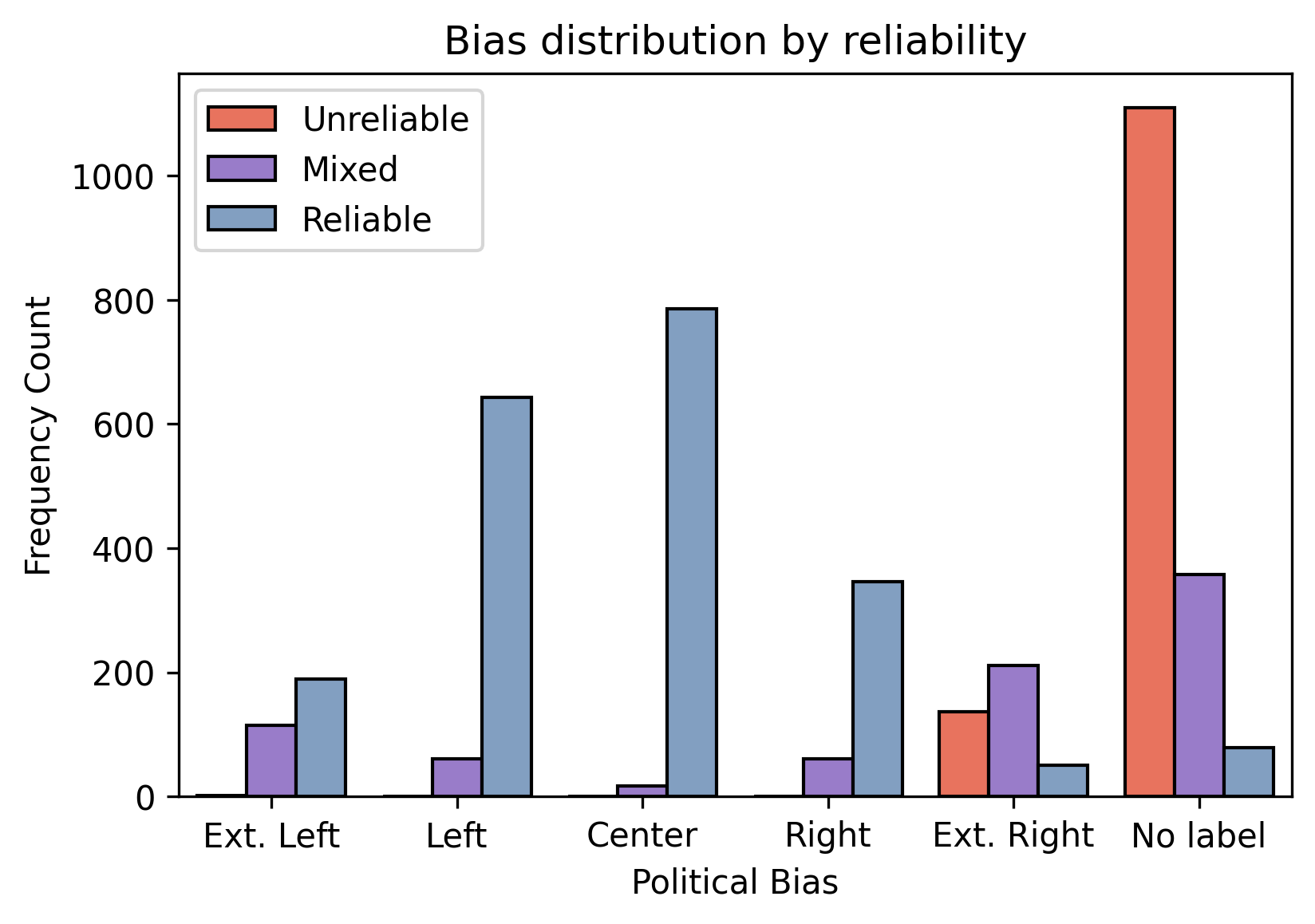}
    \caption{Bias label counts grouped by reliability}
    \label{fig:bias_by_reliability}
\end{subfigure}%
\begin{subfigure}{.47\textwidth}
    \centering
    \includegraphics[width=.95\linewidth]{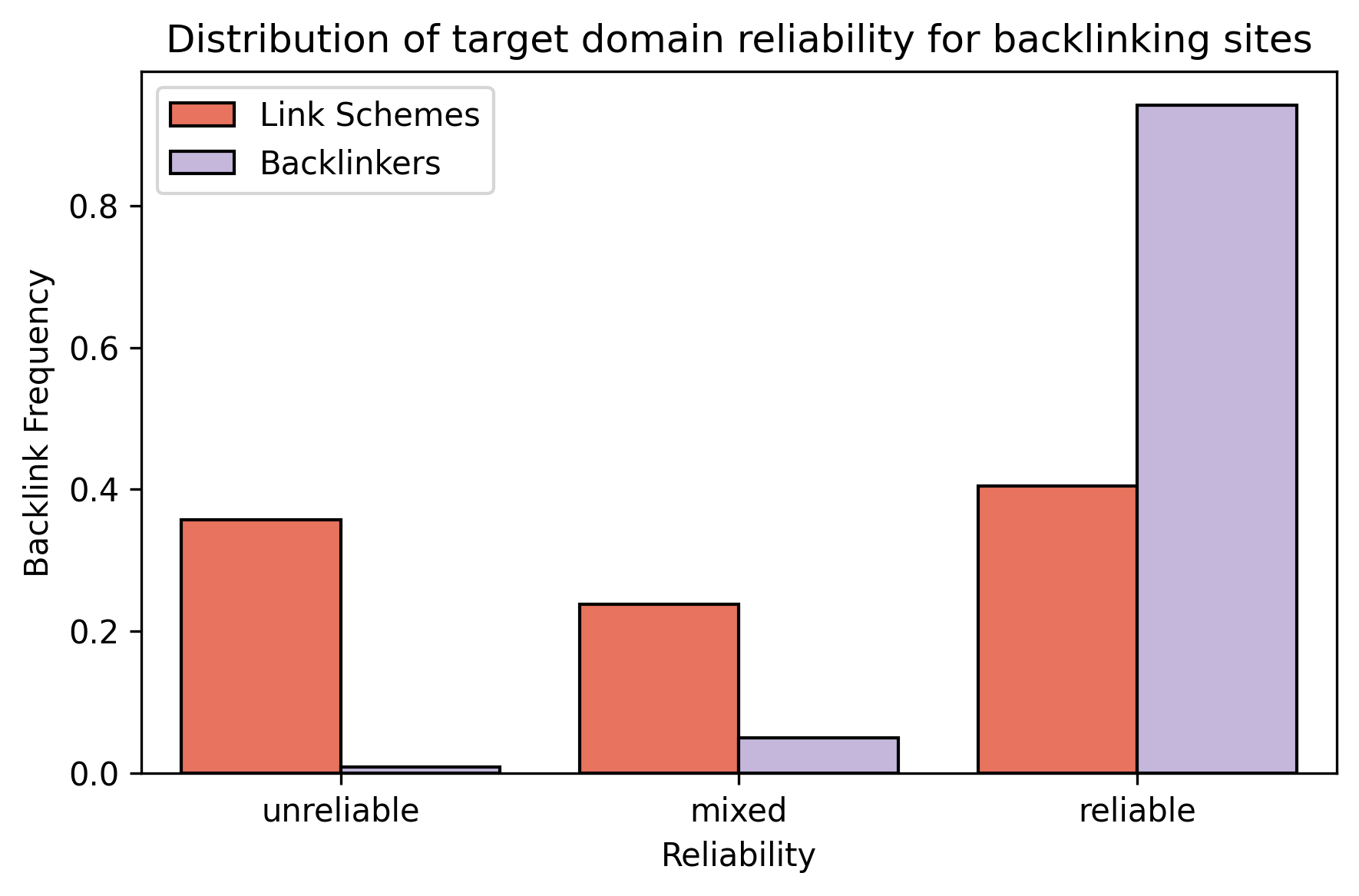}
    \caption{Link schemes vs. normal backlinking domains.}
    \label{fig:link_scheme_target_dist}
\end{subfigure}
\caption{Analysis of news domain label distributions (left) and backlinking domains (right). The distribution of link counts from source domains to targets of varying reliability reveals a significant distributional discrepancy between how link schemes and genuine backlink sites link to targets.}
\label{fig:labelled_data_and_link_schemes}
\end{figure}



A breakdown of the reliability labels according to MBFC's 2,629 political bias labels is given in \autoref{fig:bias_by_reliability}. The dataset is skewed towards reliable sites, which account for almost half the dataset (2090 of 3211 domains). \autoref{fig:bias_by_reliability} also reveals that unreliable domains with bias labels are overwhelmingly extreme right in terms of political content, motivating the principle of fairness in designing interventions. Recent collaborations between Meta and external researchers found that most misinformation existed in conservative corners of Facebook, which may suggest that this distribution reflects a reality \cite{fbideology2023}. However, we do not know how well this finding generalizes to the broader internet. Given the lack of bias labels for unreliable domains, it is challenging to design interventions that are both safe (in terms of misinformation resilience) and fair (in terms of political affiliation). Despite this challenge, the fact that unreliable labels are skewed towards the extreme right motivates the use of bias removal algorithms for unreliable interventions.


\subsection{Web Traffic Estimate Data} \label{sec:WTEData}
Aside from labels, our dataset includes 21 distinct attributes for each domain, including metrics such the total backlink and reference page counts as well as the backlink networks for each domain \citep{misinformation_detection}. We have extended our dataset to include Ahrefs traffic and ranking estimates\footnote{https://ahrefs.com/api/documentation/positions-metrics}. 

The Ahrefs `domain ranking' metric is based on their implementation of the PageRank algorithm, with scores on a logarithmic scale from 0-100. Their traffic estimate is a function of the overall search traffic of keyphrases for which the domain ranks highly and the estimated click-through rates for these keyphrases \citep{ahrefs}.

\subsubsection{Data Quality Evaluation}
Obtaining true search engine traffic numbers is not possible with public data.
To ensure data quality, we compare the Ahrefs estimates we use with commonly-used public resources. We demonstrate that Ahrefs' traffic estimates, SimilarWeb traffic estimates \citep{similarweb}, and CommonCrawl PageRank \citep{commoncrawl} are all strongly correlated.

We compare estimates for Ahrefs traffic with estimates of traffic provided by SimilarWeb \footnote{https://www.similarweb.com/}. SimilarWeb provided higher Traffic Estimates than Ahrefs for 97\% of the 2,796 domains in our dataset that it had data for. However, the rankings were very strongly correlated (Spearman $\rho= 0.855$). Log transformed traffic estimates were also strongly correlated (Pearson $r = 0.822$), and much of the variance in logged SimilarWeb traffic estimates was explainable by logged Ahrefs traffic estimates ($R^2 = 0.676$). We additionally observed strong correlation between both log-transformed Ahrefs Traffic and CommonCrawl pagerank ($r = 0.79, R^2 = 0.62)$ and log-transformed SimilarWeb Traffic and CommonCrawl pagerank ($r = 0.77, R^2 = 0.59)$

\begin{figure}
\centering
\begin{subfigure}{.45\textwidth}
    \centering
    \includegraphics[width=.95\linewidth]{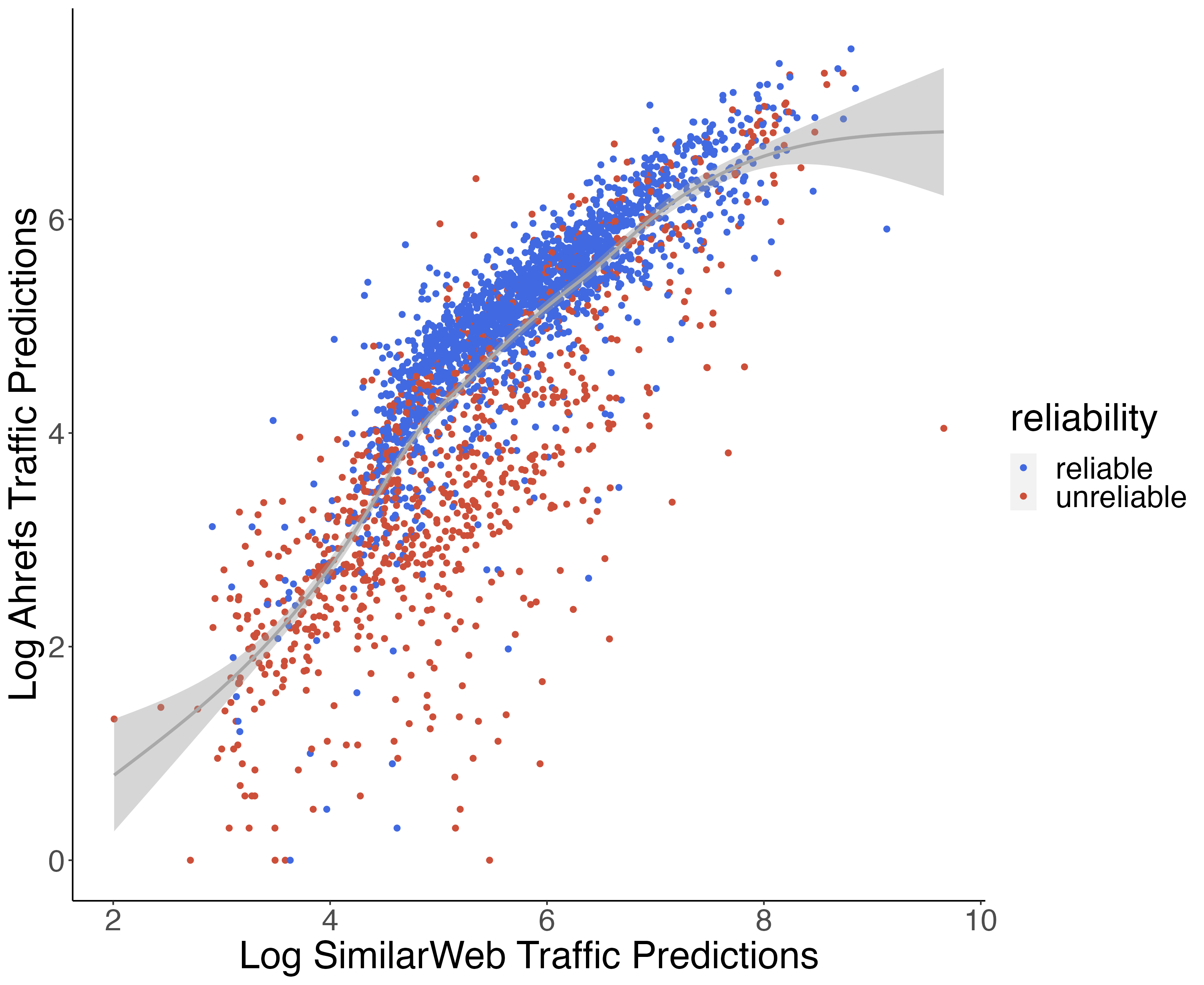}
    \caption{Ahrefs vs. SimilarWeb Log Traffic Estimates}
    \label{fig:tpc}
\end{subfigure}%
\begin{subfigure}{.45\textwidth}
    \centering
    \includegraphics[width=.95\linewidth]{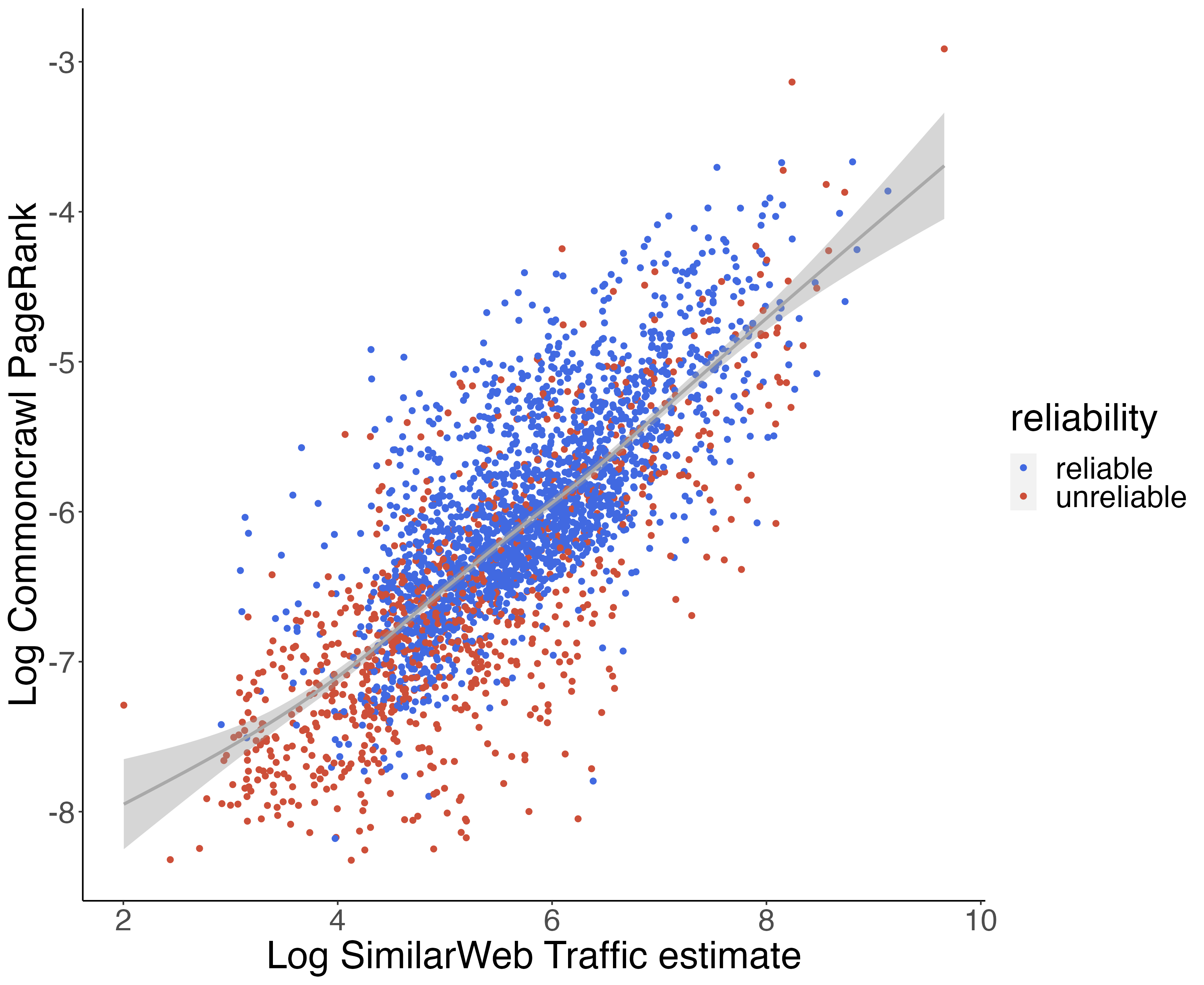}
    \caption{Common Crawl PR vs. SimilarWeb Traffic}
    \label{fig:ltc}
\end{subfigure}
\caption{Comparisons of Ahrefs Traffic Estimates and Common Crawl PageRank Estimates to SimilarWeb Estimates (logged). Blue domains are reliable by MBFC estimates and red domains are unreliable or mixed reliability.}
\label{fig:simweb_comparison}
\end{figure}

We similarly observed a strong correlation between log-transformed SimilarWeb PageRank rank Estimates and the ranks calculated from CommonCrawl data values ($\rho= 0.760$, $r = 0.760$, $R^2 = 0.577$). We visualize these traffic and PageRank Comparisons in Figure \ref{fig:simweb_comparison}.

\subsection{Webgraphs}
\paragraph{Ahrefs Backlink Network}
We utilize the backlink network from \citep{misinformation_detection}. This backlink data was originally collected from Ahrefs, taking the 10 domains that most frequently backlink to each news site. We extracted 12,351 unique backlinking domains for 3,211 unique news domains. The network is not bipartite, as news domains also sometimes link to one another. We observe a large difference between mean (2.87M) and median (47k) weighted degree, which is largely attributable to several large outlier edges. Three edges contain over 500m links: arirang.co.kr $\rightarrow$ arirang.com (1.27B), zeenews.com $\rightarrow$ india.com (0.70B), wallst.com $\rightarrow$ reuters.com (0.55B). For link scheme removal, we use a filtered version of this network that contains only those edges with a total of at least 20k backlinks and 2k referring pages. Summary statistics for this network can be found in the original paper \citep{misinformation_detection}.

\paragraph{Ahrefs Unaggregated Backlinks}
The backlink network contains 32,030 edges. Due to API limitations, we elect to drop lower-volume edges. We drop edges with fewer than 20,000 backlinks and fewer than 2,000 reference pages. This leaves us with 11,631 source-destination pairs. For each pair, we extract a sample of the 50 article-level URLs most recently visited by the Ahrefs webcrawler, where the source linked to the destination. In other words, for each pair of domains (A $\rightarrow$ B), we extract 50 articles in A that point to 50 articles in B. 


With 11,631 edges in the backlink network, each containing 50 links, this yields 576,500 links. From these we remove all URLs that end in file types (`.jpg', `.png', `.css', `.js'), leaving \urlpairs\ remaining pairs. We note that these links are not unique as a single webpage can link to the same destination multiple times, which is the case for many link scheme pages. 

\paragraph{CommonCrawl Webgraphs}
As the Ahrefs backlink network is composed of a subset of top backlinks from our labeled news domains, it acts as a testbed for the design of interventions and cannot be used to determine the unintended effects that such interventions have when propagated through the entire webgraph. 

To this end, we supplement the Ahrefs backlink network with analysis on a large-scale webgraph. The CommonCrawl foundation \citep{commoncrawl} periodically releases webgraphs based on data they scrape from the web: approximately 3 billion webpages per month \footnote{\url{https://commoncrawl.github.io/cc-crawl-statistics/plots/crawlsize}}. We use the most recently published webgraph which consists of 93.9 million domains and 1.6 billion edges between them\footnote{\url{https://data.commoncrawl.org/projects/hyperlinkgraph/cc-main-2023-may-sep-nov/index.html}}. We note that the edges in this webgraph are binarized, making interventions that rely on edge weights infeasible. Summary statistics have been published by CommonCrawl, as computed by the Webgraph Framework\footnote{\url{https://webgraph.di.unimi.it/docs/it/unimi/dsi/webgraph/Stats.html}}.

\begin{table}[!ht]
\caption{Summary Statistics for the CommonCrawl Webgraph.}
\label{tab:summary_stats}
\begin{tabular}{ll}
    \toprule
    Statistic & Backlinks \\
    \midrule
    nodes                  & 93.9M    \\
    edges                  & 1.6B    \\
    mean degree     & 17.3 \\
    largest connected component      &   37.7M nodes (40.09\%)     \\
    dangling nodes (no outlinks) & 47.7M nodes (50.8\%) \\ 
\bottomrule
\end{tabular}
\end{table}



\section{Methodology}
Our methodology follows an incremental approach. First, we seek to design and evaluate interventions on small-scale, incomplete webgraphs. Due to the incomplete nature of this data, we parameterize and fit regression models that predict the traffic and ranking of news domains. We then simulate changes to the aggregated, incomplete webgraph by removing or reweighting edges and using the previously-fit models to predict post-intervention traffic and rankings.

We then take the most successful interventions from this small-scale testbed and reimplement them on a large-scale snapshot of the webgraph. Here, we do not need to fit any models and instead carry out an analytical evaluation of webgraph interventions by computing PageRank directly. The goal here is to evaluate the interventions that had promising performance on the small-scale testbed in as close to a real-world setting as possible.

Finally, we analyze and mitigate the unintended effects of our interventions. For small-scale interventions, we explore bias removal and intervention tuning. For large-scale interventions, we investigate traffic reductions across non-news domains and refine our link scheme identification method.

\subsection{Small-Scale Webgraph Interventions}
We propose four guiding principles for the design of small-scale interventions. 1) Fairness: interventions should not penalize the link destination directly, as site owners have little control over who links to their site\footnote{With link disavowal being an exception: \url{https://support.google.com/webmasters/answer/2648487}}, 2) Generality: interventions should not be aware of link destination reliability labels, 3) Cost: interventions should increase the cost or complexity of adversarial behavior, and 4) Scalability: interventions should have the potential for web-scale implementation.

\subsubsection{Link Scheme Removal Intervention}
Previous work has found that black-hat SEO methods are used disproportionately in favor of unreliable news sources \citep{misinformation_detection, williams2023search, link_scheme_misinfo, bradshaw2019disinformation}. The motivation for this intervention is that link-scheme sites should not be capable of affecting news domain rankings. Typically, farmed links have little or no context (e.g. a list of links with no content), and exist purely to game search rankings. As such, we employ a heuristic approach to detecting link schemes based on their link distributions. We define a link scheme as a domain through which the site owner creates a huge number of links, on the order of 100,000, to increase a target site's search ranking and by extension its traffic.

We identify the link scheme domains on which we run interventions using the identification algorithm proposed in \citep{misinformation_detection}. The algorithm relies on finding a small number of link schemes sites that heavily link to known unreliable domains, with criteria governing the requirements for the minimum number of backlinks going to unreliable domains (depth $\alpha$), and the minimum number of unreliable domains linked to (breadth $\beta$).

This method was originally proposed to detect unknown unreliable sources, where the full set of available unreliable labels are used, with $\alpha = 3$ and $\beta = 400000$ \citep{misinformation_detection}. However, using unreliability labels to identify link schemes violates the principle of generality. To mitigate the loss of generality, we identify link scheme sites based on a 20\% random sample of our unreliable labels. Since backlinking sites will have fewer backlinks pointing to a 20\% sample than to the full list of unreliable domains, we lower the original link scheme identification parameters, using $\alpha = 2$ and $\beta = 100,000$. We aim to demonstrate that this intervention is effective given limited knowledge of the destination domains, and publicly release the list of 250 link scheme sites identified by our algorithm \citep{misinformation_detection}.

Williams and Carley list several structural elements such as the lack of `about' sections or paid advertising spaces that are common among link schemes and are independent of misinformation labels \citep{williams2023search}. While such labels are used for convenience here, future work on link scheme identification that does not rely on misinformation labels is warranted. 

\subsubsection{Link Multiplicity Intervention}

The Link Multiplicity intervention is aimed at decreasing the effectiveness of popular link-farming techniques by re-weighting edges in the backlink network using a set of link scores. One example of the link-farming method that we are attempting to downrank is the same URLs being embedded in many identical sidebars across a backlink site. The motivation for this intervention follows from the analysis of link scheme outlinks, which tend to disproportionately link to unreliable sites \autoref{fig:link_scheme_target_dist}.

We begin with the set of \urlpairs links in our unaggregated Ahrefs dataset. We define multiplicity over (`source domain'$\rightarrow$`destination url') pairs as the number of links to the destination URL that occur over all pages of the source domain. A high level of multiplicity may indicate that the source domain is engaging in link spam on several of its pages. Alternatively, the links may embedded in a structural part of the site, such as on a sidebar that gets rendered on every page under the given source domain.

We then invert link multiplicity frequencies to downrank links with high multiplicity, before applying minmax scaling to generate multiplicity scores for each (`source domain'$\rightarrow$`destination url') pair. The resulting scores are in the range of [0,2], so links with high multiplicity are penalized with a low score, $<1$, and links with low multiplicity are boosted with a high score, $>1$. Finally, we aggregate the scores by taking the mean, leaving a single score for each edge in the backlink network.

\subsubsection{Tuning the Link Multiplicity Intervention} 
Reliable news sites are of crucial importance for the dissemination of accurate reporting. Regardless of the effect that these interventions may have on misinformation sites, it is natural that a search engine provider would be hesitant to introduce any such intervention that would adversely impact the traffic that reliable sites attract. We demonstrate that our methods may be tuned to reduce impact, and may be used to promote sites rather than downrank them.

Link multiplicity intervention scores are generated using a Minmax scaler ($\mathrm{min} = 0, \mathrm{max} = 2$). As the multiplicity scores are used in the regression model to weight the backlink attribute, the Minmax parameters can be modified to shift the distribution of the transformed scores such that they downrank unreliable sites less, but promote reliable sites more.


\subsubsection{Debiasing the Interventions}
Given the predictive nature of the small-scale webgraph models, they may learn to predict reliability based on political leaning. We note that the large-scale webgraph methods do not have this limitation as they are analytic and not model-based. On the small-scale webgraph, we propose the use of algorithmic debiasing methods for mitigating the potential for the regression models to learn a political bias based on label imbalance. This is important as MBFC domain-reliability rankings contain more unreliable right-leaning domains than unreliable left-leaning domains \autoref{fig:bias_by_reliability}. 

To mitigate the impacts of politically imbalanced starting seeds, we employ a Disparate Impact Remover \citep{disparate_impact} to pre-process our dataset before the regression experiments so that we can compare the performance of interventions using regression models trained on the original SEO attributes and the debiased SEO attributes\footnote{Alternative debiasing methodologies such as in-processing (which mutates learning objectives) and post-processing (which mutates output predictions) are more challenging to integrate with the models and evaluation metrics used here.}. 

The disparate impact level of a dataset is defined in terms of the ratio of the probabilities of favorable labels for "unprivileged" and "privileged" groups (\autoref{eq:disparate_impact}). As such, it ranges between 0 and 1, with 1 indicating equality of outcome. In this case, favorable implies the reliable label, and privilege is associated with the bias label $B$. The "privileged" group, based on the bias of domain reliability labels, has extreme left bias and the "unprivileged" group has extreme right bias as in \autoref{eq:disparate_impact}. We explore disparate impact across various levels of "repair" $R$, which modifies the underlying data to protect a given class. $R=0$ returns the original dataset and $R=1$ is "fully repaired" \cite{feldman2015certifying}.

\begin{equation}
    DI = \frac{P(\textrm{reliable} | B = \textrm{Extreme Right})}{P(\textrm{reliable} | B = \textrm{Extreme Left})}
    \label{eq:disparate_impact}
\end{equation}



First, we train a random forest classifier, to classify domains based on their SEO attributes as reliable or unreliable, using binarized reliability labels where "mixed" reliability is treated as unreliable. We train the classifier with $n=50$ estimators. This allows us to measure fairness by the disparate impact level in the predicted outcomes of the classifier. We then debias the dataset by applying a disparate impact remover \citep{aif360}, and verify that the predictions of the random forest model have an improved disparate impact score when trained on the debiased attributes. As such, we measure fairness in terms of outcomes of extreme right domains vs extreme left domains, and more specifically by the ability of the reliability classifier to not label all extreme right-wing domains as unreliable.

We then run the same interventions as before on the debiased dataset. We recompute intervention evaluation metrics on the results using a new set of regressors, fit to the debiased dataset. The motivation is to create search ranking systems that are both safe from misinformation and fair in terms of political bias. However, we note that there is often a tradeoff between algorithmic performance and fairness. In this case, given the underlying correlation between political bias and reliability labels (\autoref{fig:bias_by_reliability}), it is unlikely that algorithmic bias can be fully mitigated.


\subsection{Small-Scale Webgraph Evaluation Metrics}
We evaluate the success of these interventions based on their ability to simultaneously penalize unreliable sites and maintain the existing domain ranking and traffic levels of legitimate and reliable news sources. Consequently, we group the predicted change in traffic and ranking by reliability. If unreliable domains are heavily penalized and reliable domains are not, the intervention is predicted to be successful. To calculate the predicted change given an incomplete webgraph, we fit regression models to predict traffic and ranking. We then run the intervention and use the models to predict post-intervention traffic and rankings on the modified webgraph. Given the incompleteness of the small-scale webgraph and the predictive nature of the models, this method acts as a testbed to evaluate candidate interventions before implementing them for the large-scale CommonCrawl webgraph \citep{commoncrawl}.

\subsubsection{Model Selection}
We consider several different models to predict traffic and rank---a linear regression, a vanilla 2-layer neural network, and two graph neural networks, one which incorporates a binarized backlink network ($GNN_{uw}$) and one which incorporates a weighted backlink network ($GNN_{w}$). We compare MSE on log-traffic and log-pagerank prediction for each, and model details and specifications are provided in Appendix \ref{app:NB}. We find that the unweighted GNN model outperforms the weighted GNN model, the vanilla neural network, and the regression model on both the log-traffic and log-PageRank rank prediction tasks (\autoref{tab:model_comparisons}). While the regression performs worse than the unweighted graph neural network and the vanilla neural network on both prediction tasks, its MSE is still relatively low. We determine that this slight decrease in MSE is worth the interpretability of the regression coefficients and therefore elect to use linear regressions over GNNs in evaluation. We leave the tradeoff between simple attribute-based models and more complex GNNs as an avenue for future work and refer the reader to to \cite{misinformation_detection} for a deeper discussion on the matter.

\begin{table}[]
\caption{Model 10-run test MSE average and standard deviation for Ahrefs Traffic and CommonCrawl PageRank prediction. The Neural Network (NN) and Regression models use only Ahrefs features of labeled domains. The unweighted $GNN_{uw}$ and log-weighted $GNN_{w}$ use these features along with the backlink network extracted from Ahrefs.}
\label{tab:model_comparisons}
\begin{tabular}{lcc}
\toprule
           & \multicolumn{1}{l}{\textbf{Traffic MSE}} & \multicolumn{1}{l}{\textbf{PageRank MSE}} \\
\midrule
$GNN_{uw}$    & $3.932 \pm 0.15$                    & $0.81 \pm 0.04$                      \\
$GNN_{w}$     & $7.35 \pm 2.86$                     & $2.86 \pm 0.31$                      \\
linear     & $4.27 \pm 1.5$                      & $0.87 \pm 0.15$                      \\
regression & 5.07                            & 4.85                            \\
\bottomrule
\end{tabular}

\end{table}

\subsubsection{Analysing the Regression Model}
To better understand the data, we perform regression analysis on all 21 attributes that we pull from Ahrefs. Of these 24 attributes, 10 are highly correlated with Pearson coefficients $> 0.9$. Of the remaining 11 attributes, 6 are significant at the $p=0.05$ threshold for the domain ranking regression. We run our log-transformed regressions on the uncorrelated and significant attributes for the entire dataset as well as for each of the unreliable (-unrel), mixed, and reliable (-rel) label groups independently, as shown in \autoref{tab:regression_full}. Notably, $R^2$ is high for each dependent variable and for each category of reliability label, and the majority of variables are significant at the $p=0.01$ level for both tasks.

Additionally, the backlink attribute has one of the highest coefficients for the traffic regression and the highest for the rank regression. We therefore expect that our backlink-based interventions will have an impact on the dependent variables in both the traffic and domain ranking regressions. We note that for both dependent variables, the backlink coefficient is larger for reliable domains than for unreliable domains. In fact, for the traffic regression on the unreliable domain dataset, the backlink coefficient is close to zero. This indicates that naively removing backlinks will have a substantially worse impact on reliable domains than on unreliable reliability domains. This necessitates a targeted approach to webgraph-based interventions, where indiscriminately removing backlinks is likely to do more harm than good.

In order, the other domain attributes from \autoref{tab:regression_full} are; HTML pages---the total number of webpages under the given domain; External links---the number of links leading away from the domain; Internal links---the number of self-referential links; User-generated content (UGC)---the number of links that come from off-site content such as discussion forums and comment sections on blogs; Redirect---the number of other domains that redirect to this domain; Canonical (root)---number of links that point to the canonical or root version of the domain, without specifying a specific webpage or article; EDU---the number of links that come from .edu domains.

\begin{table}
\caption{Baseline regression results for prediction of Traffic (top) and Rank (bottom) on different label sets.}
\label{tab:regression_full}
\begin{tabular}{llllllllll}
\toprule
   \textbf{DV}   &   \textbf{R$^2$} &  \textbf{Backlink}    & \textbf{Pages} & \textbf{Ext.} & \textbf{Int.} & \textbf{UGC} & \textbf{Redirect}    & \textbf{Root} &  \textbf{EDU}  \\
\midrule
Traffic &     0.94                  & 0.19** & 0.06 & -0.1** & 0.22** & -0.12** & 0.17** & 0.26** & 0.51** \\
 -unrel & 0.83                       & 0.04 & 0.21** & -0.1** & 0.12** & 0.09 & 0.14** & 0.2** & 0.43**           \\
 -mixed & 0.92                       & 0.42** & 0.21* & -0.28** & 0.18* & -0.3** & 0.09 & 0.23** & 0.48**           \\
 -rel   & 0.97                       & 0.36** & 0.03 & -0.2** & 0.35** & 0.11** & 0.12** & 0.17** & 0.12**          \\
 \midrule
 Rank   &       0.97                 & 0.31** & -0.09** & 0.02* & 0.09** & -0.01 & -0.05** & -0.04** & -0.05** \\
 -unrel & 0.94                       & 0.29** & -0.02 & -0.02 & 0.06** & 0.05** & -0.05** & -0.06** & -0.04**           \\
 -mixed & 0.97                       & 0.28** & -0.17** & 0.06** & 0.13** & 0.0 & -0.03 & -0.04** & -0.07**            \\
 -rel   & 0.98                       & 0.35** & -0.09** & 0.03** & 0.07** & -0.07** & -0.06** & -0.03** & -0.05**      \\
 \bottomrule
 \end{tabular}
 \end{table}

\subsubsection{Simplified Regression Model for Intervention Evaluation}
Of the 8 attributes in the traffic and rank regressions (\autoref{tab:regression_full}), we select three variables of interest that will enable us to evaluate our interventions fairly.
\paragraph{Backlinks $(X_1)$:} the total number of links that point to each of our destination (labeled news) domains. This gives an approximation of the link network that is outside the control of the destination domain. The backlink attribute will be mutated as we selectively negate and weigh backlinks by source reputation for each intervention, so it is included by necessity.
\paragraph{External links $(X_2)$:} the total number of outbound links from the destination domain, encompassing the part of a domain's link network that is within its control. According to our principle of fairness, this attribute will not be mutated, and as such acts as a counterbalance to the backlink mutation.
\paragraph{HTML pages $(X_3)$:} the total number of webpages under the given domain. This gives an idea of a domain's footprint in terms of total content and acts as another counterbalance so as not to overestimate the impact of our interventions. As opposed to the external links attribute, the total number of HTML pages of a domain is not intrinsically related to the domain's webgraph. We include this variable to help control for website size within the regression.

\subsubsection{Traffic and Rank Reduction}
We evaluate small-scale interventions using regressions trained to predict organic traffic and domain ranking. We predict log traffic and log ranking using the simple linear log-transformed regression model in \autoref{eq:regression}.

\begin{equation}
ln(y) = \beta_0 + \beta_1 ln(X_1) + \beta_2 ln(X_2) + \beta_3 ln(X_3)
\label{eq:regression}
\end{equation}


We fit traffic and ranking regression models and then mutate the dataset to simulate the effects of introducing each of the interventions in turn. That is, we remove and downrank backlinking sites by mutating the relevant SEO attributes for each domain. More formally, for an intervention that applies penalty $\delta$ to $X1$ (\# backlinks), we examine \autoref{eq:regression_delta}.

\begin{equation}
    ln(y^*) = \beta_0 + \beta_1 ln(X_1(1-\delta)) + \beta_2 ln(X_2) + \beta_3 ln(X_3) 
\label{eq:regression_delta}
\end{equation}

We use these regression models to predict the change in traffic and ranking, the \% retrained (RP), and average this over all domains with a given reliability label D as in \autoref{eq:reduction}.

\begin{equation}
RP(D) = \frac {1}{|D|}\sum_{d}^{D} \frac{y_d^*}{y_d} 
\label{eq:reduction}
\end{equation}

Finally, we report the \textit{Reliability Impact Score} (RIS), a convenience statistic that facilitates the comparison of interventions. This rewards the reduction of traffic and domain ranking for unreliable and mixed reliability domains while penalizing reductions for reliable domains. It is simply the difference between the percentage of rank and traffic that is retained between each of the reliability categories, according to the regression models as in \autoref{eq:RIS}.

\begin{equation}
RIS =  RP(\mathrm{reliable}) - \frac{RP(\mathrm{unreliable}) + RP(\mathrm{mixed})}{2}
\label{eq:RIS}
\end{equation}

The RIS metric treats mixed and unreliable labels equally. A score of 0 indicates that reliable and unreliable domains have equal reductions in the dependent variable (traffic or rank). A score of 0.1 means that we have reduced predicted traffic to unreliable and mixed news domains by 10\% more than for reliable domains. A higher RIS metric is therefore more desirable.


\subsection{Large-Scale Webgraph Interventions}
We adapt the link scheme removal intervention from our small-scale webgraph (32k edges) and evaluate it on a large-scale snapshot of the webgraph with 1.6 billion edges. We present an analytical evaluation by computing PageRank directly on CommonCrawl data \citep{commoncrawl}, emulating a real-world ranking system.

\subsubsection{Link Scheme Identification}
As the CommonCrawl webgraph is binarized, we employ a simplified version of the link scheme identification algorithm \citep{misinformation_detection} that ignores link counts ($\alpha_{min}$) and relies only on the number of unreliable domains a given backlink site links to ($\beta_{min}$). As the ATR algorithm is sensitive to the initial list of link spam domains used, we conservatively set the $\beta_{min}$ threshold to 200. We implement the ATR algorithm by running PPR on the reversed edge list (i.e., edge $i\rightarrow j$ becomes $j \rightarrow i$). 

We run this algorithm on the set of identified link schemes to produce an extended seed list for our interventions which we call the ATR Extended List. We expect that the edge removal and Inverse-PPR interventions will have an increased impact when run with the ATR Extended List as opposed to the link scheme list, as the ATR Extended list should contain more link scheme domains that crosslink to the original link scheme list \citep{jiyoung_whang_scalable_2020}. 

Aside from the simplified version of the link scheme identification algorithm \citep{misinformation_detection} and the use of the ATR algorithm to extend the resulting list of link schemes, the link scheme removal intervention is unchanged from the small-scale webgraph experiments---we simply remove all links coming from this list of link scheme domains. We forgo implementing the multiplicity intervention as it relies on edges weighted by link counts and so is incompatible with binarized webgraphs.

\subsubsection{Inverse PPR} 
We hypothesize that while the PPR algorithm is well suited to promoting certain sites, it will have greater unintended effects on domain ranking across the webgraph. We develop the notion of Inverse PPR for deranking the preference vector instead of upranking it. Specifically, we initialize the PageRank vector to be 0 for all domains in the preference vector, with uniform probability for all other domains. As such, we hypothesize the Inverse-PPR method will influence PageRank away from domains in the preference vector without the unintended effects that come with running PPR. The Inverse-PPR method is to be contrasted with the Edge Removal intervention, which we expect will have a more pronounced effect. We run the Inverse-PPR and Edge Removal interventions for both the seed list of link scheme domains and the ATR expanded list.

\subsubsection{Analyzing Unintended Effects}
We use large-scale webgraphs to identify the unintended effects of our interventions in a real-world search engine setting. To this end, we investigate the distribution of changes in CommonCrawl PageRank as a result of our interventions for all domains in the webgraph, not just news domains. We then take the domains whose PR score changes significantly (\>50\%) and group these domains based on their domain names and SimilarWeb domain categories. We use these groupings to categorize the types of domains that utilize link spam SEO services and those that are affected by our interventions. 

\subsubsection{Multi-Category Link Scheme Identification}
Whang et. al. demonstrate that the ATR algorithm is sensitive to the initial list of seed domains \citep{jiyoung_whang_scalable_2020}. With this in mind, we propose a method for improving the precision of initial lists. Our method for filtering out non-link schemes from the initial seed list relies on the context-free nature of link scheme SEO providers. These SEO providers place links on their sites for a price, and as such have a myriad of URLs across various categories; e-commerce, gambling, news, health, sports, etc. Although there are some notable exceptions (i.e. the online encyclopedia wikipedia.org\footnote{Which normally contains only no-follow links, but this distinction is not observed in public CommonCrawl dataset.}), this behavior is mostly unique to multi-category link schemes. As a result, sites that link to multiple low-quality domains across various categories are highly likely to be multi-category link schemes.

It has previously been observed that scam casinos and sports betting sites employ black-hat SEO methods \citep{yang_casino_2019, phillips_search_2024}. We use a curated blocklist of scam casino and sports-betting sites\footnote{\MYhref{https://www.casino.org/untrustworthy/}{This list} is based on a \MYhref{https://www.casino.org/how-we-rate/}{thorough review process} that rates online casinos and sports betting sites by factors such as site security measures, licensing, regulatory oversight, and payment processes.} to identify multi-category link schemes by taking the intersection between domains that have a high ATR score for both the scam casino and unreliable news lists.

\subsection{Large-Scale Webgraph Evaluation Metrics}
To validate that the effectiveness of experiments designed on small-scale, purpose-built webgraphs translates to large-scale, noisy webgraphs, we replicate the link scheme removal intervention on the latest CommonCrawl webgraph release \citep{commoncrawl}. This allows us to evaluate our interventions by computing PageRank directly, instead of using a regression model to predict it. That is, we determine the effectiveness of the intervention based on how much it reduces the PageRank of unreliable domains. It also allows us to quantify the unintended effects of running such interventions by its effect on any domain in the webgraph, as opposed to just news domains.

To compute PageRank over the CommonCrawl webgraph, we utilize the Webgraph Framework \citep{boldi_webgraph_2004}. This framework allows us to evaluate our interventions against the Personalized PageRank (PPR) algorithm \citep{jeh_scaling_nodate} and the Anti TrustRank (ATR) algorithm \citep{jiyoung_whang_scalable_2020}\footnote{We implement these within the open-source \MYhref{https://github.com/vigna/webgraph}{Webgraph Framework}. We provide \MYhref{https://github.com/PeterCarragher/cc-webgraph}{the source-code} for running all experiments, which builds upon open-source utilities \MYhref{https://github.com/commoncrawl/cc-webgraph}{provided by CommonCrawl}}.


\section{Small-Scale Webgraph Results}

\subsection{Traffic Regression}
We run regressions on Ahrefs attributes extracted from the labeled (destination) news domain data. The goal is to predict traffic and domain rank as functions of backlink counts. As larger sites may naturally have larger backlink volumes, we control for domain size using the total number of HTML Pages that the domain contains as well as the number of its external links. 

We display the results for the log-transformed regression, i.e., where a log transformation is applied to the dependent as well as to each independent variable, in \autoref{tab:regression_base}. We can interpret the coefficients as the approximate percentage change in a dependent variable resulting from a percent change in an independent variable \cite{benoit2011linear}---a 1\% increase in backlinks is associated with a 0.422\% increase in rank and 0.276\% increase in traffic, all else held constant. The impact of HTML pages and External Links appears more mixed. A 1\% increase in HTML pages is associated with a 0.586\% increase in rank, but a -0.081\% decrease in traffic. Finally, a 1\% increase in external links is associated with a -0.1387 decrease in rank, but a 0.067\% in traffic. We find that these simple models explain a large proportion of the variance in traffic and domain rank with adjusted R$^2$ of 0.907 and 0.960 respectively. We report results for our evaluation method on the pre-intervention Ahrefs dataset in \autoref{tab:regression_base}. Results on the debiased dataset are discussed in full detail in Section \ref{sec:BR}.

\begin{center}
\begin{table}
\caption{Baseline regression results on the three variable model per dependent variable.}
\label{tab:regression_base}
\begin{tabular}{lllll}
\toprule
  \textbf{Dependent Variable}   &   \textbf{R$^2$} &  \textbf{Backlinks}    & \textbf{HTML Pages} & \textbf{Links External} \\
\midrule
Log(Traffic)      &     0.907 & 0.422** & 0.586** & -0.139**\\
Log(Traffic) Debiased & 0.892 & 0.719** & -0.338** & 0.323** \\
Log(Rank)       &       0.960 & 0.276** &  -0.081** & 0.067** \\
Log(Rank) Debiased & 0.973 & 0.262** & -0.132** & 0.116** \\
\bottomrule
\end{tabular}
\end{table}
\end{center}


 %

\subsection{Link Multiplicity Intervention}

\begin{figure}
\centering
\begin{subfigure}{.5\textwidth}
  \centering
    \includegraphics[width=.95\linewidth]{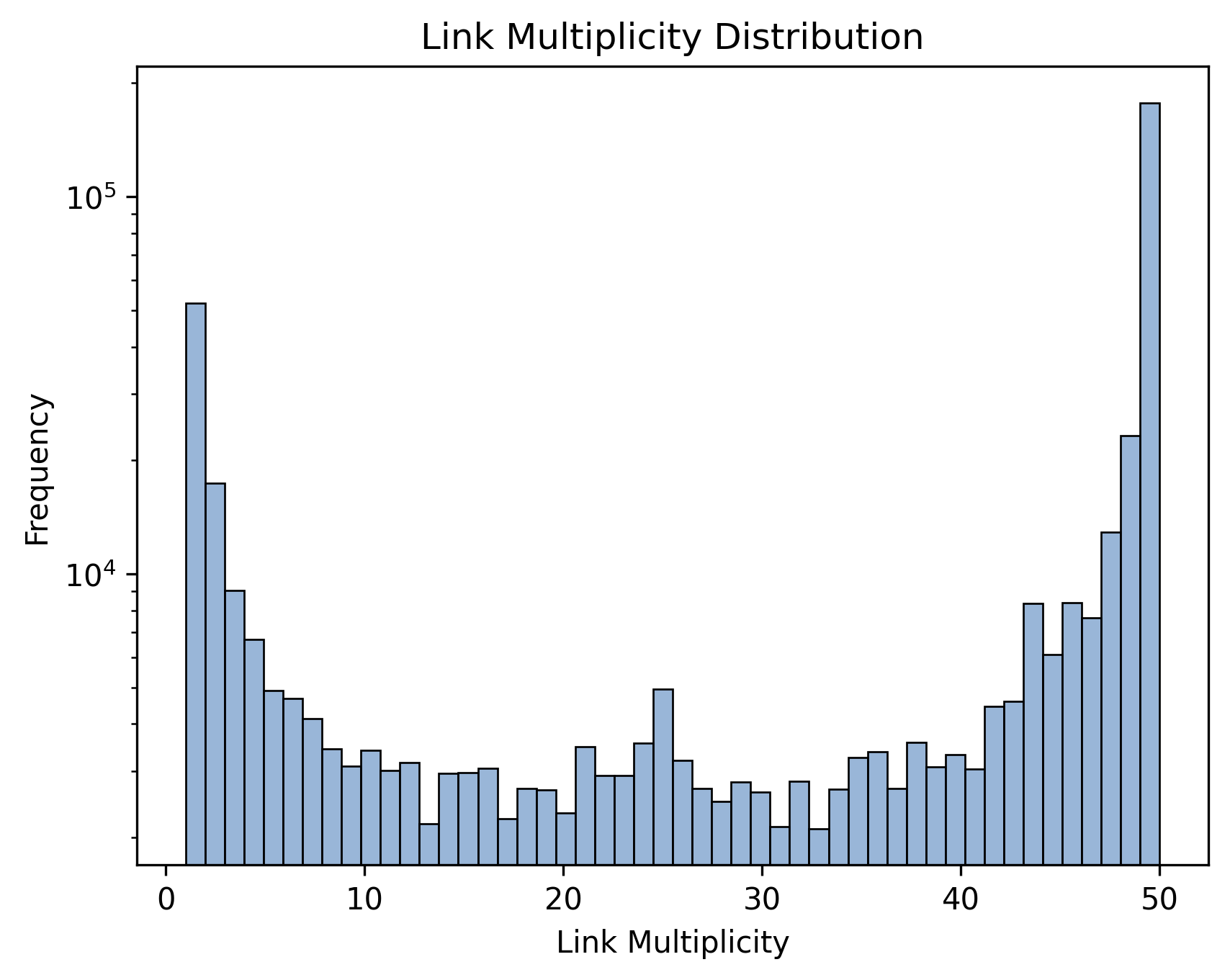}
    \caption{Link Multiplicity distribution.}
    \label{fig:multiplicity_dist}
\end{subfigure}%
\begin{subfigure}{.5\textwidth}
  \centering
  \includegraphics[width=.95\linewidth]{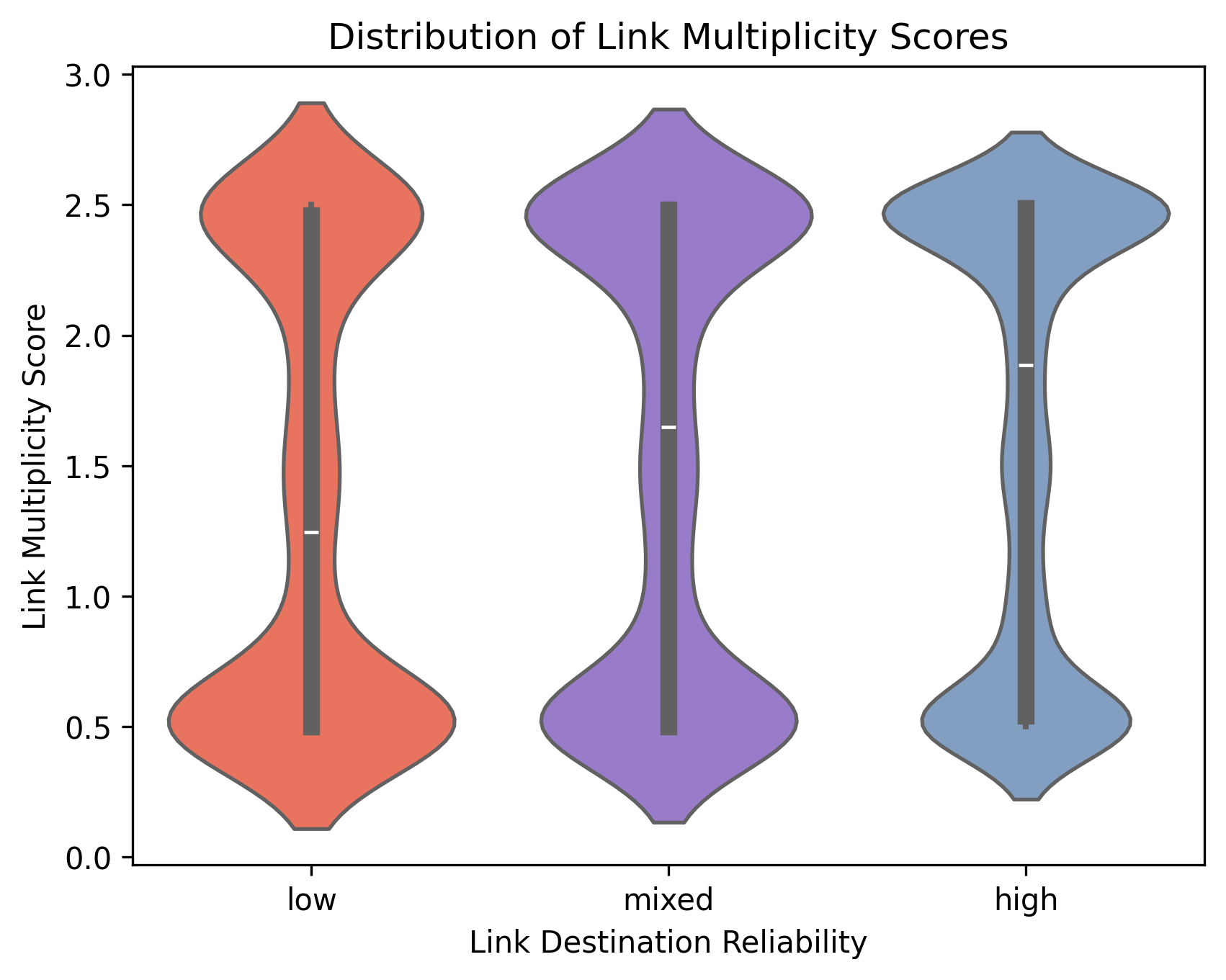}
  \caption{Inverted and min-max scaled multiplicity scores.}
  \label{fig:multiplicity_violin}
\end{subfigure}
\caption{The link multiplicity distribution (left) reveals a dichotomy between links that appear a handful of times and links that occur extremely frequently on a source domain ($>40$ times). Link Multiplicity scores (right) are derived from this link multiplicity distribution and capture this dichotomy with a distinct dumbbell-like shape.}
\label{fig:multiplicity}
\end{figure}

We profile the multiplicity of links within the \urlpairs\ URL pairs that we pull from the Ahrefs backlink API. We find that links follow a bimodal distribution---either a link appears once or twice within the source domain, or it appears 50 times. Note that as a result of API limitations, we pull a maximum of 50 links per edge in the backlink network, making 50 an upper bound. The distribution of link multiplicity for (source domain, destination URL) pairs is provided in \autoref{fig:multiplicity_dist}.

Inverted link multiplicity scores are assigned to each edge pair in the backlink network, penalizing backlink edges that have high multiplicity with a score $<1$ and boosting backlink edges that have low multiplicity with a score $>1$. These scores are then grouped by destination domain reliability, and the resulting distributions are given in \autoref{fig:multiplicity_violin}. Despite similar bimodal distributions across each reliability group (arising from \autoref{fig:multiplicity_dist}), there here is a larger skew towards zero for low-reliability domains, and towards two for high-reliability domains, meaning that unreliable domains tend to have more backlinkers with $>50$ links to the same page (the scores are inverted). As these scores are used in the intervention to scale edge weights, link multiplicity scores should on average penalize unreliable domains more than reliable domains.

\subsection{Evaluation} 
\autoref{tab:intervention_traffic_regressions} details the results of our intervention evaluation metric which gives the proportion of destination domain rank and traffic remaining in the post-intervention prediction. In other words, it is the post-intervention prediction as the numerator with the pre-intervention traffic prediction as the denominator. The control intervention shows that simply removing a certain percentage of backlinks from each destination domain does not favor any reliability category in particular. This further highlights the importance of targeted interventions. 

In our small-scale webgraph setting, we find that the link scheme removal intervention is successful at achieving its objective. The intervention reduces the traffic and rank of unreliable and mixed domains threefold compared to its impact on unreliable domains---achieving a 35\% reduction in unreliable traffic at the expense of an 11\% reduction in reliable traffic. In a real-world setting, an 11\% traffic reduction to reliable sites would likely not be considered acceptable by search engine companies, but we note that this number can be tuned while the ratio is maintained. The intervention yields a percent traffic reduction to unreliable domains that is $>$3x larger than that of reliable domains. The Multiplicity intervention is also successful, but less so than the Link Scheme intervention, with a RIS of around 0.05 (\autoref{tab:intervention_traffic_regressions}).

Notably, combining interventions leads to an improvement in RIS (\autoref{tab:intervention_traffic_regressions}). The combination of Link Scheme and Multiplicity interventions has a summative effect on traffic reductions, with the decrease in traffic and rank from the Link Scheme intervention offset by increases from the multiplicity scores. This yields a 32\% reduction for unreliable traffic at the lower cost of a 5\% reduction for reliable traffic. As the goal here is to drop unreliable traffic while maintaining reliable traffic, the combined intervention presents a more measured approach to this tradeoff.

\begin{table*}[]
\caption{Results showing the proportion of estimated traffic and domain rank remaining after each intervention. A three-variable regression on backlinks, html\_page, and external\_link attributes is used.}
\label{tab:intervention_traffic_regressions}
\begin{tabular}{lllllllll}
\toprule
Intervention & \multicolumn{4}{c}{\makecell{Traffic}} & \multicolumn{4}{c}{\makecell{Rank}} \\
\cmidrule(lr){2-9}
                            & Unreliable & Mixed & Reliable & RIS & Unreliable & Mixed & Reliable & RIS\\
\midrule

\textbf{L}ink Scheme  & 0.65 & 0.77 & 0.89 & 0.18       & 0.73 & 0.82 & 0.92 & 0.14      \\
\textbf{M}ultiplicity & 1.01 & 1.04 & 1.07 & 0.05       & 1.00 & 1.02 & 1.04 & 0.03      \\
L+M Combined          & 0.68 & 0.80 & 0.95 & 0.21       & 0.75 & 0.83 & 0.95 & 0.16      \\
\midrule
Control 0\%           & 0.00 & 0.00 & 0.00 & -0.00      & 0.04 & 0.03 & 0.03 & -0.01     \\
Control 50\%          & 0.72 & 0.72 & 0.72 & 0.00       & 0.83 & 0.83 & 0.83 & 0.00      \\
Control 100\%         & 1.00 & 1.00 & 1.00 & 0.00       & 1.00 & 1.00 & 1.00 & 0.00     \\
\bottomrule
\end{tabular}
\end{table*}


\section{Unintended Effect Mitigation in Small-Scale Webgraphs}

In this section, we explore the potential limitations and unintended consequences of our proposed interventions on the small-scale network and explore different mitigation strategies. We explore boosting reliable domains rather than penalizing unreliable domains and attempt to mitigate the issue of political bias present in our seed list.

\subsection{Tuning Small-Scale Interventions to Reduce Impact on Reliable News Sites}

We first demonstrate that the multiplicity intervention can be modified to boost reliable sites rather than penalize unreliable sites. The multiplicity intervention can be tuned such that the estimated remaining traffic of reliable sites is not impacted (i.e. $\mathrm{remaining traffic} = 1$). Scores above 1 indicate that these interventions can potentially be used to boost reliable sites, instead of focusing on penalizing unreliable and mixed domains. We provide illustrative results for a series of multiplicity intervention tuning experiments in \autoref{fig:tuning_multplicity}, demonstrating the effects of the intervention in promoting reliable sites over unreliable sites are maintained regardless of how it is tuned. To mitigate the impact of the link scheme removal intervention, future work could modify the link scheme identification algorithm \citep{misinformation_detection} to identify reliable backlinking sites instead of link scheme sites.



\subsection{Bias-Removal in Small-Scale Models}\label{sec:BR}
\begin{figure}
\centering
\begin{subfigure}{.33\textwidth}
  \centering
    \includegraphics[width=.95\linewidth]{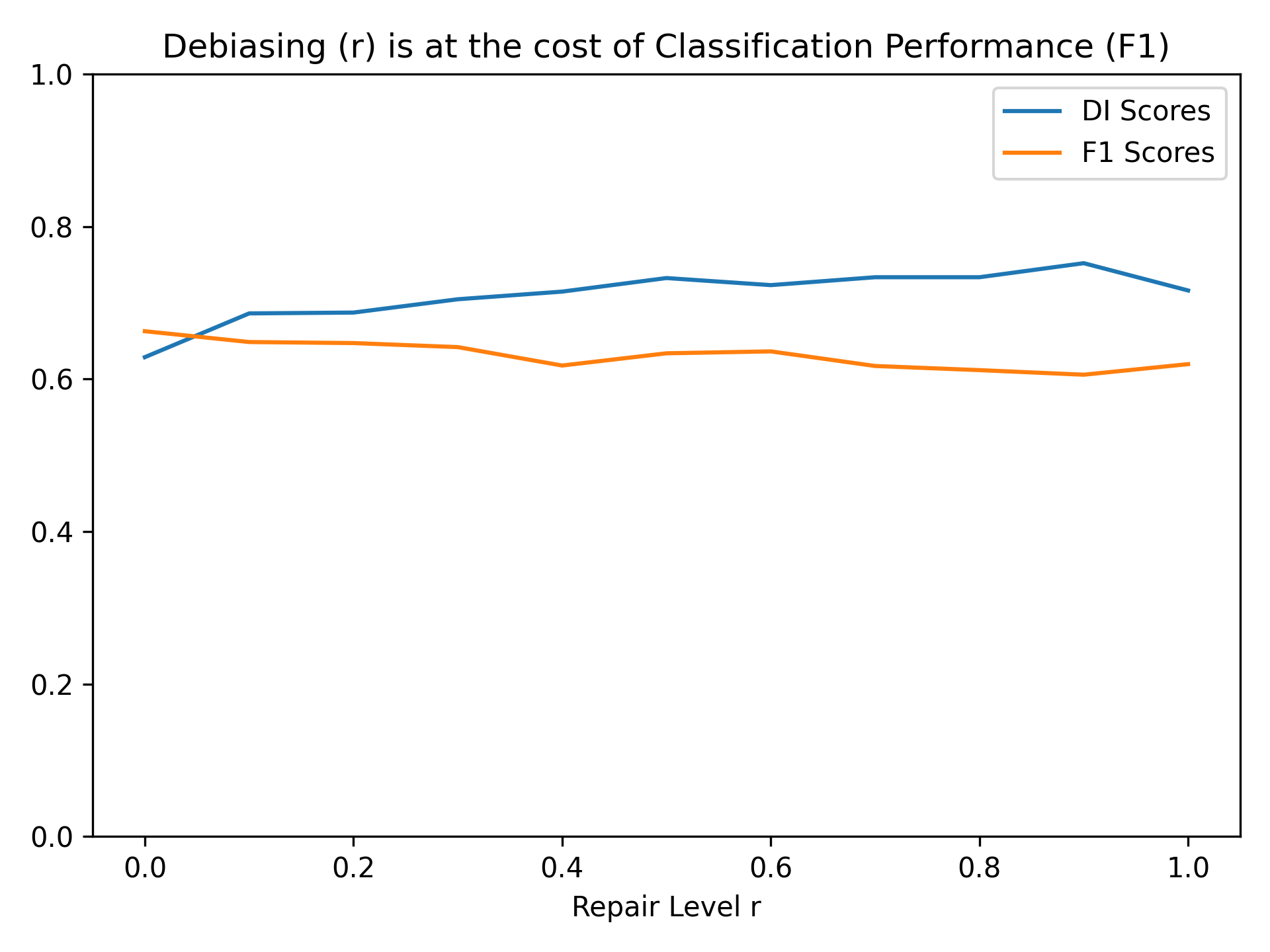}
    \caption{Classification Performance}
    \label{fig:disparate_impact_vs_perf}
\end{subfigure}%
\begin{subfigure}{.33\textwidth}
  \centering
  \includegraphics[width=.95\linewidth]{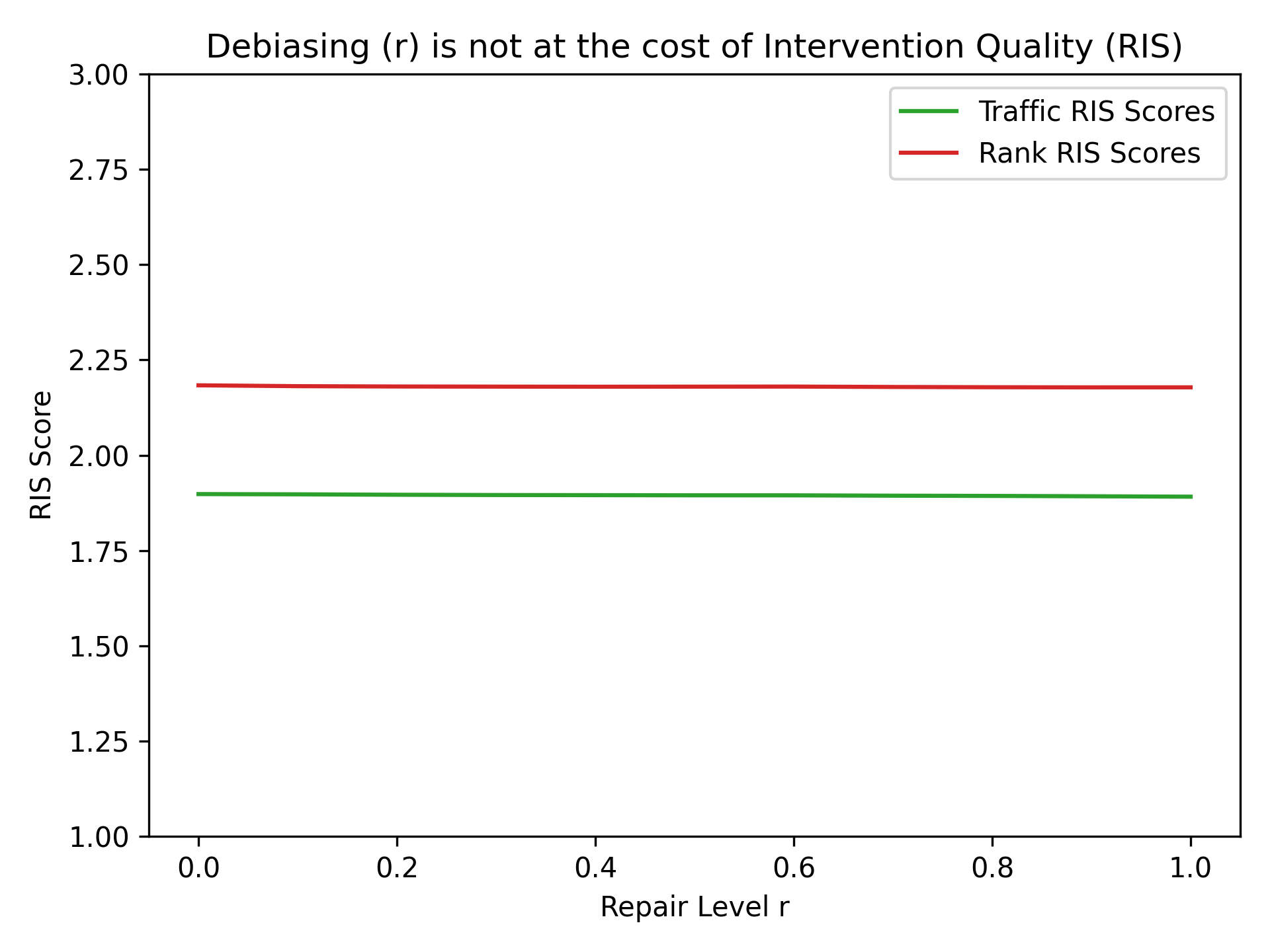}
  \caption{Intervention Quality}
  \label{fig:ris_scores}
\end{subfigure}
\begin{subfigure}{.33\textwidth}
    \centering
    \includegraphics[height=3.5cm]{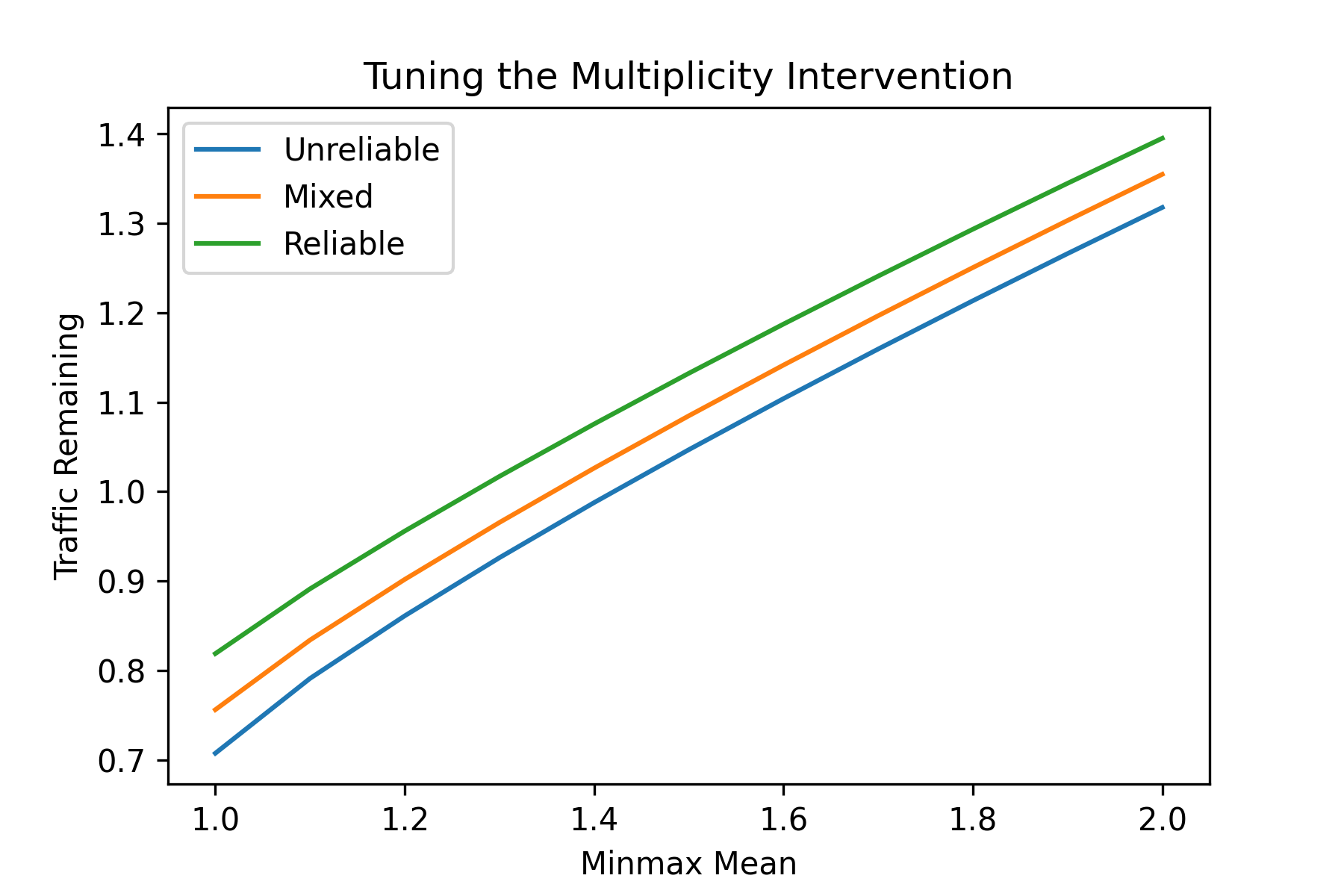}
    \caption{Tuning Multiplicity}
    \label{fig:tuning_multplicity}
\end{subfigure}
\caption{Experiments in mitigating unintended effects of large-scale interventions. As we debias the dataset, performance on reliability classification tasks slightly degrades but intervention quality mostly stays constant. As repair level R increases, disparate impact (DI) slightly improves alongside a slight decline in F1 score on the reliability classification task (left). However, RIS is constant for the link scheme removal intervention across all repair levels during debiasing (center).}
\label{fig:debias_results}
\end{figure}

We evaluate our small-scale interventions on a subset of the Ahrefs websites that contain MBFC political bias labels. Disparate impact removal is parameterized by the repair level $R$, where 0 is the original data and 1 is "fully repaired" as defined in \cite{feldman2015certifying}. We show the impact that varying $R$ has on both the level of disparate impact and the performance of a news domain reliability classifier. \autoref{fig:disparate_impact_vs_perf} demonstrates that as we increase repair level, disparate impact improves (increases) meaning that algorithmic bias learned from our dataset decreases, according to political leaning. However, performance on the classification task declines implying that debiasing could affect intervention performance.

To test this theory, we take our best-performing intervention, link scheme removal, and evaluate the quality of this intervention using Reliability Impact Scores as we increase the repair level $R$ of disparate impact removal. For each repair level $R$, disparate impact removal returns a new set of attributes on which we run our intervention and generate the RIS metric. The result is a trend that shows intervention quality remains constant throughout the debiasing procedure, regardless of the repair level used (\autoref{fig:ris_scores}). This is in contrast to our expectations, and it signals that interventions do not have to come at the cost of fairness, where fairness concerns the political leaning of news domains. Additionally, \autoref{tab:regression_base} shows that debiasing the dataset does not adversely impact regression quality.

A complete set of intervention results for the debiased dataset is given in \autoref{tab:intervention_traffic_regressions_bias}. We find that interventions on debiased SEO attributes are equivalent to those run on the original attributes when predicting rank, and even slightly improve performance when predicting traffic with a 0.02 increase in RIS. This finding complements \autoref{tab:intervention_traffic_regressions} and provides further evidence that debiasing the dataset has a non-negative impact on intervention quality (\autoref{tab:intervention_traffic_regressions_bias}). 

\begin{table*}[]
\caption{Results showing the proportion of estimated traffic and domain rank remaining after each intervention. Interventions are split on original or debiased attributes (repair level $R=1$) and are run on a dataset filtered to contain only those news domains for which bias scores are available.}
\label{tab:intervention_traffic_regressions_bias}
\begin{tabular}{lllllllll}
\toprule
Intervention & \multicolumn{4}{c}{\makecell{Traffic}} & \multicolumn{4}{c}{\makecell{Rank}} \\
\cmidrule(lr){2-9}
                            & Unreliable & Mixed & Reliable & RIS & Unreliable & Mixed & Reliable & RIS\\
\midrule
Original Attributes &&&&&&&&\\
\midrule
\textbf{L}ink Scheme  & 0.81 & 0.79 & 0.90 & 0.10      & 0.88 & 0.87 & 0.94 & 0.07      \\
\textbf{M}ultiplicity & 1.03 & 1.06 & 1.10 & 0.05      & 1.00 & 1.02 & 1.04 & 0.02      \\
L+M Combined          & 0.84 & 0.83 & 0.99 & 0.16      & 0.89 & 0.88 & 0.97 & 0.09      \\

\midrule
Debiased Attributes &&&&&&&&\\
\midrule

\textbf{L}ink Scheme  & 0.78 & 0.76 & 0.89 & \textbf{0.12}      & 0.88 & 0.87 & 0.94 & 0.07      \\
\textbf{M}ultiplicity & 1.05 & 1.08 & 1.13 & \textbf{0.07}      & 1.00 & 1.02 & 1.04 & 0.02      \\
    L+M Combined          & 0.83 & 0.81 & 1.00 & \textbf{0.18}      & 0.89 & 0.88 & 0.97 & 0.09  \\

\bottomrule
\end{tabular}
\end{table*}

\section{Large-Scale Webgraph Results}
Given the strong performance of the link scheme intervention on the small-scale webgraph, we now evaluate it against several well-known baseline methods in a large-scale search ranking experiment, and extend the link scheme list using ATR scoring to enhance its effectiveness.

\begin{table}[]
    \centering
    \caption{Intervention results on the CommonCrawl webgraph are evaluated by PR centrality and PR rank as computed using the Webgraph Framework \citep{boldi_webgraph_2004}.}
    \label{tab:cc_webgraph_intervention_results}
    \begin{tabular}{lllllllll}
\toprule
Intervention & \multicolumn{4}{c}{\makecell{PageRank Centrality}} & \multicolumn{4}{c}{\makecell{PageRank Rank}} \\
\cmidrule(lr){2-9}
                            & Unrel & Mixed & Reliable & RIS & Unrel & Mixed & Reliable & RIS\\
\midrule
PPR on Reliable News               & 4.731 & 4.903 & 646 & 642 & 5.103 & 4.277 & 203.119 & 198 \\
Inv-PPR on Unreliable News         & 0.621 & 0.660 & 1.000   & 0.359   & 0.642 & 0.765 & 1.000   & 0.297   \\
\midrule
Inv-PPR on Link Schemes            & 0.999 & 1.000 & 1.000   & 0.001   & 0.998 & 1.000 & 1.000   & 0.001   \\
Inv-PPR on Anti-TrustRank          & 0.951 & 0.984 & 0.995   & 0.027   & 0.940 & 0.982 & 0.995   & 0.034   \\
\midrule
ER on Link Schemes       & 0.917 & 0.944 & 0.970   & 0.040   & 0.899 & 0.876 & 0.947   & 0.060   \\
ER on Anti-TrustRank     & 0.897 & 0.931 & 0.976   & 0.062   & 0.878 & 0.887 & 0.947   & 0.064   \\
ER on Multi-Category ATR & 0.951 & 0.957 & 0.989   & 0.035   & 0.963 & 0.948 & 0.960   & 0.004 \\

\bottomrule
    \end{tabular}

\end{table}

\subsection{PPR (Reliable) and Inv-PPR (Unreliable) Baselines}
\autoref{tab:cc_webgraph_intervention_results} presents results for the PPR and Inv-PPR baseline interventions when we ignore the principle of fairness and directly intervene on our labeled news domain---they represent upper bounds when these interventions are taken to the extreme, with Inv-PPR on unreliable labels achieving the highest RIS metric of 0.359 across both small-scale and large scale experiments. Although applying PPR to promote reliable domains appears to perform exceptionally well according to the RIS metric (642), this method has more profound unintended effects, with both PageRank Rank and PageRank Centrality increasing by 500\% for unreliable news domains. This is due to the average deviation of the preference vector from the uniform distribution being orders of magnitude higher for PPR than for Inverse-PPR. For Inverse-PPR only link scheme domains have a significant deviation from the uniform distribution, whereas, for PPR, every domain is changed significantly. Note that PageRank centrality is here used in place of traffic estimates, as we can compute it directly and have previously shown that it correlates strongly with Ahrefs traffic estimates (\autoref{fig:simweb_comparison}).

\subsection{Link Scheme and Anti-TrustRank Interventions}
We find that Inv-PPR under-performs on the RIS metric compared to Edge Removal for both link scheme and ATR Extended lists. Furthermore, applying ATR to extend domain lists before running interventions is beneficial, with the ATR Extended domain list out-performing the link scheme list from which it is derived on both Inv-PPR and Edge Removal interventions. We discuss the multi-category intervention which aims to reduce the unintended effects of Edge Removal in \autoref{sec:multi_category_intervention}.

While edge removal interventions have the desired effect ($RIS > 0$), we find rank is better retained when interventions are run on the large-scale webgraph (0.062) than the small-scale web graph (0.18). As the CommonCrawl webgraph is binarized, these results serve as lower bounds for intervention effectiveness when link counts are not considered and increased webgraph context introduces noise to PageRank calculations. Conversely, the small-scale Ahrefs webgraph acts as an upper bound.

\begin{figure}
\centering
    \includegraphics[width=.9\linewidth]{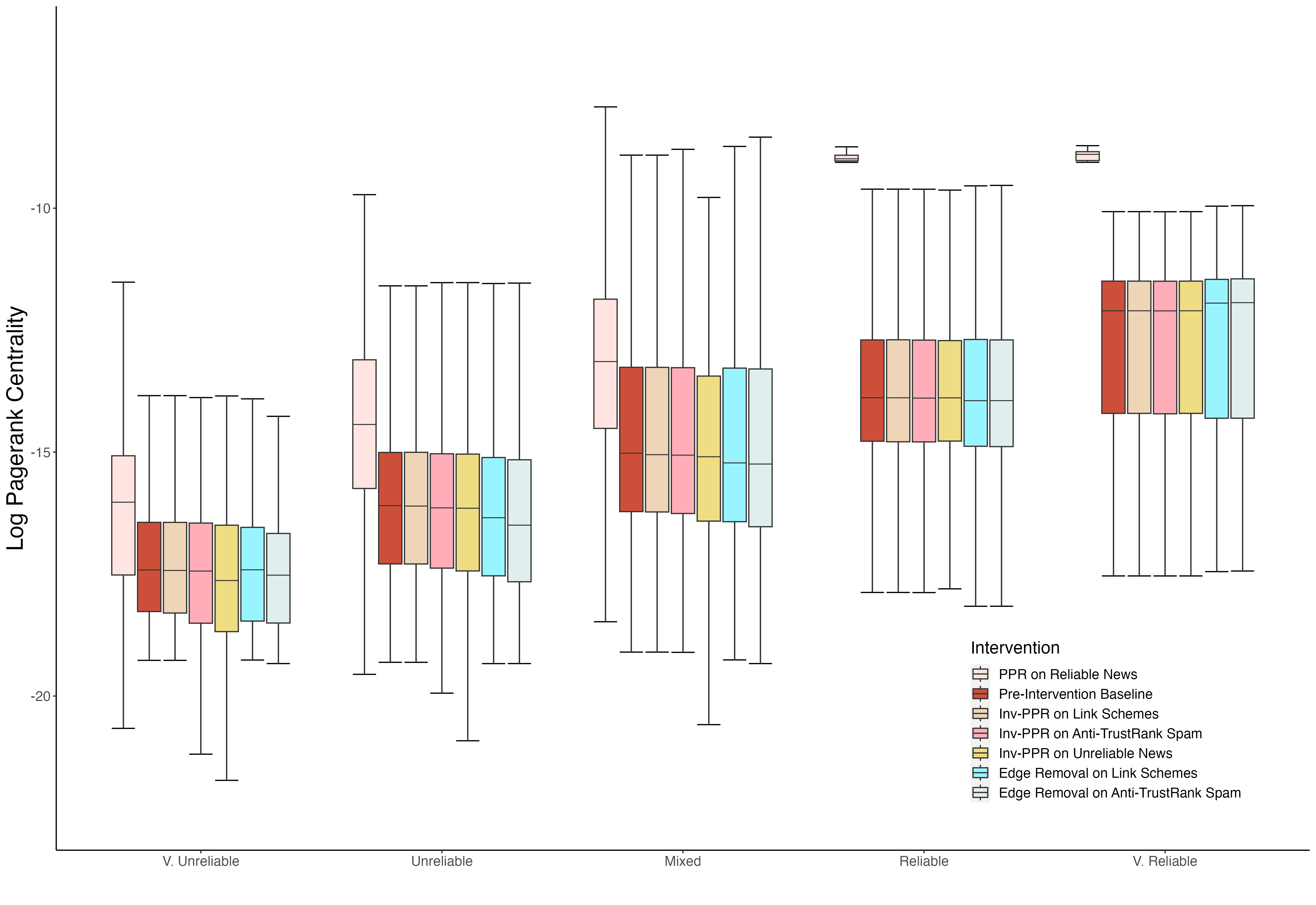}
    \caption{PR changes per intervention, grouped by reliability. Lower-reliability news sites are more affected than higher-reliability sites, which remain consistent.}
\label{fig:cc_webgraph_intervention_dist}
\end{figure}

The distribution of PR changes that occur as a result of removing outlinks from link schemes and Anti-TrustRank sites is further broken down by reliability label in \autoref{fig:cc_webgraph_intervention_dist}. We find statistically significant decreases in PR centrality for lower reliability domains, with limited impact on reliable domains as demonstrated in table \autoref{tab:cc_webgraph_intervention_results}---the \% retention of PR centrality is $> 97\%$ for reliable domains on all interventions.



\section{Unintended Effect Mitigation in Large-Scale Webgraphs}
The motivation for implementing interventions on the large-scale webgraph is two-fold---demonstrating that these interventions can be applied in a real-life setting without relying on predictive models and exploring the unintended effects that such interventions can have. Unlike in small-scale webgraphs, we can assess classes of websites that are inadvertently affected by large-scale interventions. Taking the ATR Extended Edge Removal intervention as our best-performing "fair" intervention, we first investigate the performance of the intervention in terms of its accuracy in identifying link schemes. This is necessary as the ATR algorithm is known to be sensitive to the initial seed list \citep{jiyoung_whang_scalable_2020} and so inaccuracies in link scheme identification will lead to larger unintended effects. Next, we explore the extent to which the intervention adversely impacts domains outside of our labeled news dataset. Finally, we develop the notion of Multi-Category Anti-TrustRank (MC-ATR) as a means to mitigate these unintended effects by improving the accuracy of link scheme identification.

\subsection{ATR Extended Edge Removal Analysis}
We observe that many domains in the ATR extended list are link-scheme sites. Of 10 domains with the highest ATR scores, six form a single SEO service network (articlement.com, heavenarticle.com, globalseoarticles.com, webranksite.com, top2high.com, kingranks.com\footnote{\url{https://kingranks.com/guest-post-add-my-site/}}), two are blogging platforms popular among link schemes (blogspot.com, wordpress.com), one is a separate SEO service (webseodirectory.com), and only one is legitimate (wikipedia.org). This supports the claim that when seeded with link scheme domains, domains that rank highly according to their Anti-TrustRank score are new link scheme domains that link heavily to the original link scheme list \citep{jiyoung_whang_scalable_2020}. 

Despite this finding, removing link schemes from the webgraph affects all domains that link schemes link to, not just unreliable news domains. To assuage these concerns, we find that after the edge removal intervention on the ATR Extended list, the majority of domains are relatively unaffected---84.7M out of 93.9M domains have a change in PR centrality that is less than 5\%. However, 2.8M domains experience a PR centrality reduction of over 10\% (with 0.7M increasing by more than this), and 1.3M experience a reduction of more than 20\% (with 100k sites increasing by more than this). \autoref{fig:pr_reduction_dist} in the appendix shows the complete distribution of PR centrality changes. 

To determine the types of domains where unintended effects are felt most strongly, we analyze all domains whose original PageRank is at least 1e-7 and whose reduction in PR centrality due to the intervention is at least 50\%. For the ATR extended edge removal intervention, these are 7k domains, 3.2k of which SimilarWeb lists domain categories for \citep{similarweb}. Here we find that categories prevalent among heavily affected domains include alternative health sites (16), gambling (14), and sports-betting (4). This analysis supports previous work that interrogates the use of link scheme services by scam casinos and sports betting sites \citep{yang_casino_2019, phillips_search_2024}.  


However, we note that this approach also negatively impacted many "university and college" (89) and "law and government" domains (172), suggesting that link scheme sites are engaging in camouflage by co-citing certain classes of reliable domains, but also establishing the need for additional algorithmic protections for governmental and university domains. Unsurprisingly, News \& Media (293) is the single most affected category. Within it, we observe unreliable news sources that are not in the original labeled dataset, but also some seemingly-reliable local news sources. These observations motivate important avenues for future research.

\begin{figure}
\centering
\begin{subfigure}{.45\textwidth}
    \centering
    \includegraphics[width=.9\linewidth]{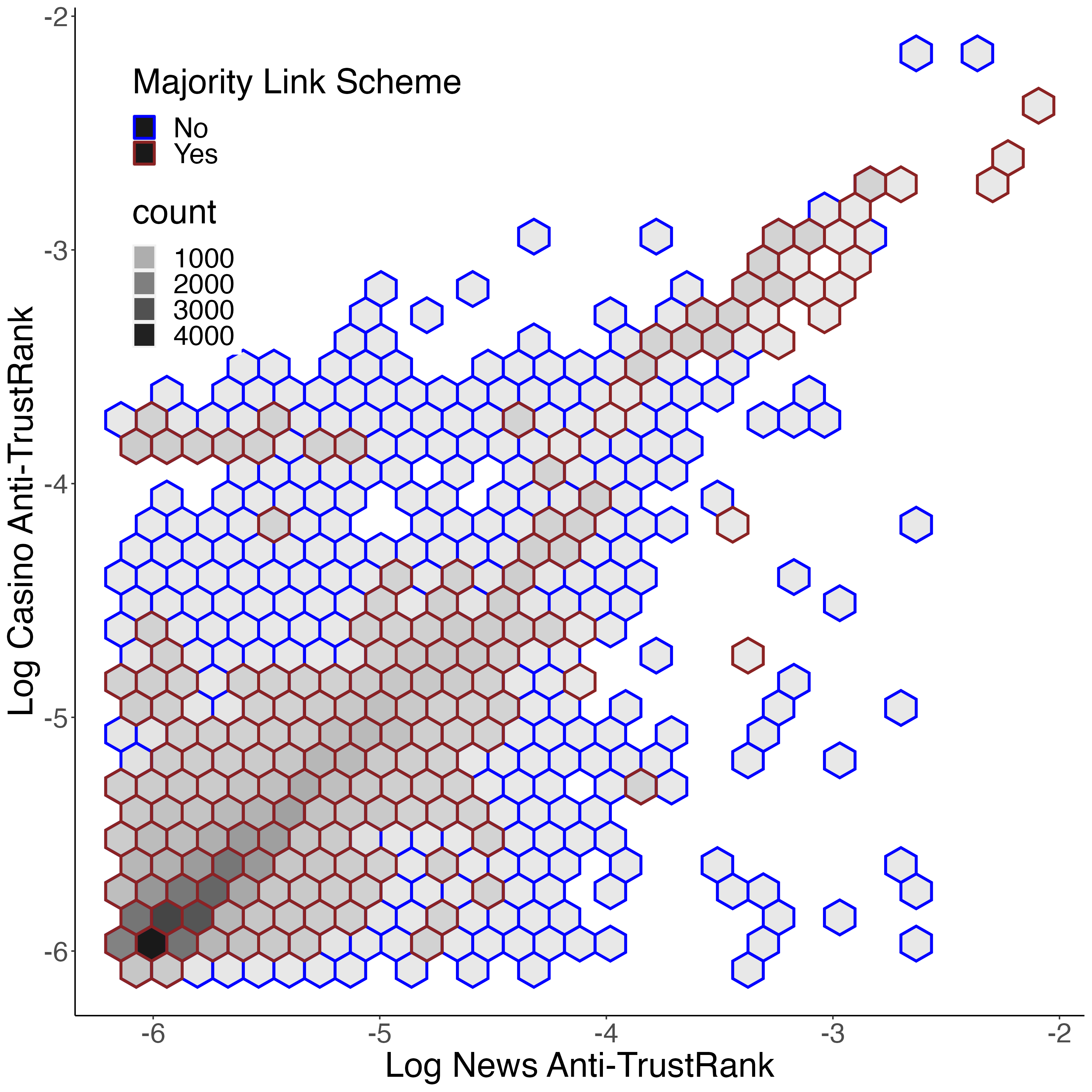}
    \caption{Multi-Domain Link Scheme Identification}
    \label{fig:multi-domain-identification}
\end{subfigure}%
\begin{subfigure}{.55\textwidth}
    \centering
    \includegraphics[width=.95\linewidth]{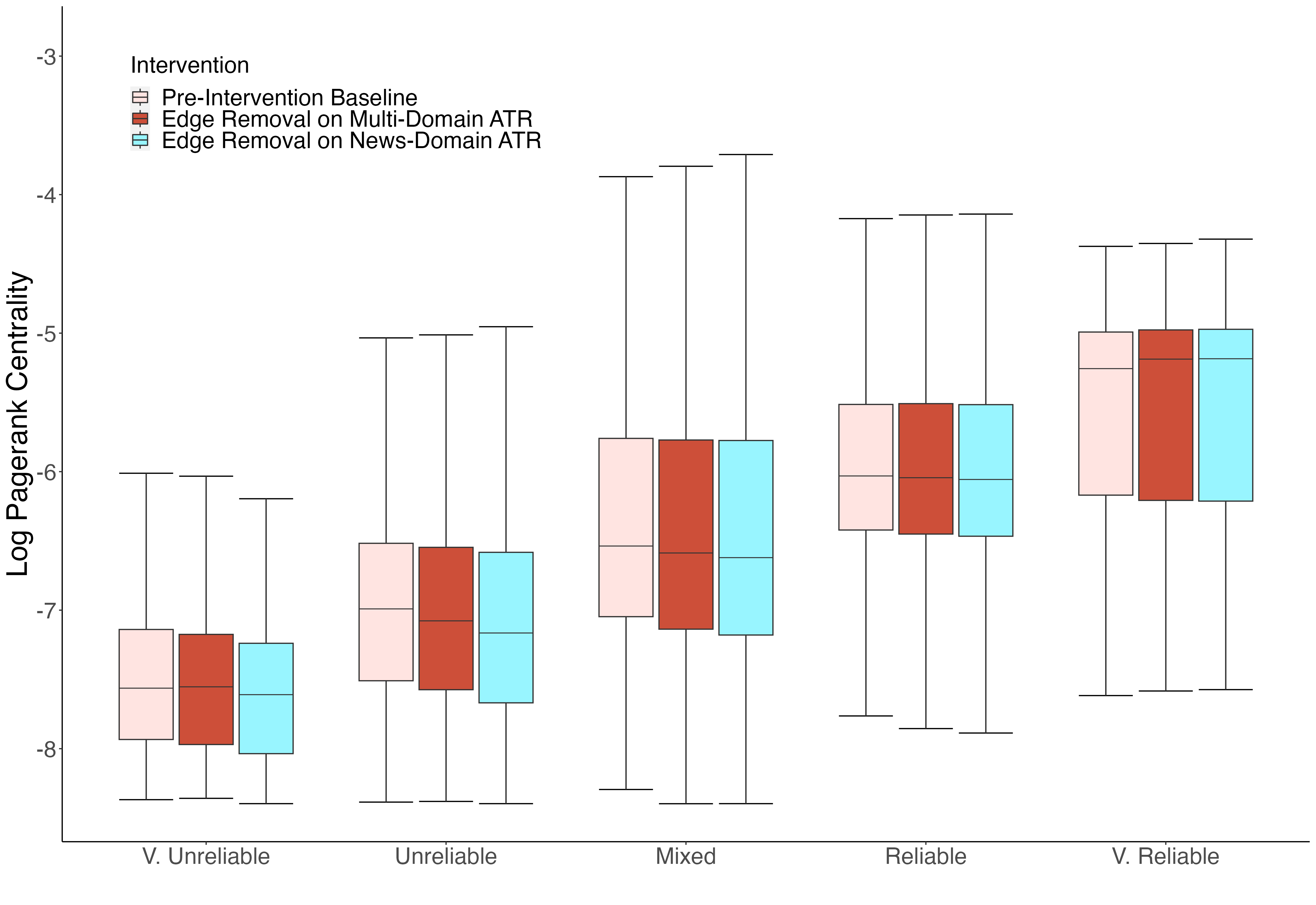}
    \caption{Multi-category Intervention effectiveness using}
\label{fig:cc_webgraph_intervention_dist_multi_domain}
\end{subfigure}
\caption{Left: Sites that link to both unreliable news and scam casino sites as given by the ATR algorithm tend to be link spam providers. Right: restricting the seed list to multi-domain link spam providers reduces the impact of intervention effectiveness.}
\label{fig:cc_webgraph_multi_domain}
\end{figure}

\subsection{Multi-Category Anti-TrustRank} \label{sec:multi_category_intervention}
Inspired by this finding, and in an attempt to increase the accuracy of both the link scheme identification algorithm and the interventions that rely on it, we propose a definition of multi-category link scheme services as domains that serve to inauthentically increase traffic to target sites of arbitrary type. As a result, multi-category link scheme providers link to target domains of entirely unrelated categories, creating a unique linking pattern that we can take advantage of in the link scheme identification process (albeit at the cost of recall, as shown in\autoref{tab:cc_webgraph_intervention_results} and \autoref{fig:cc_webgraph_intervention_dist_multi_domain}).

To demonstrate this, we take the list of scam casino domains and run the ATR algorithm to find sites that link heavily to them. We then take domains that have ATR scores $>10^{-4}$ for both the unreliable casino and news site lists---we call these multi-category link schemes. We plot the full list of multi-category link schemes in  (\autoref{fig:multi-domain-identification}). We color bins red if the majority of domains within the bin contain words that are heavily associated with link-scheme websites\footnote{URLs containing any of the substrings: `SEO', `directory', `rank', `link', `article', or `site'}. Of the 70 sites with the highest ATR scores, 65 were link schemes. Of these, 25 are part of a single link scheme network, the services of which can be purchased through Fiverr \footnote{\url{https://www.fiverr.com/bilalarticles?source=gig_page}}. 


Removing the top 1K domains in terms of ATR score on this multi-category seed list has the following effects on the edge removal intervention: relative to the ATR edge removal intervention, using the multi-category domain list resulted in 1M (or 10\%) fewer domains having their rank changed by more than 5\%. However, this comes at the cost of reducing the effectiveness of the intervention (\autoref{fig:cc_webgraph_intervention_dist_multi_domain}). As such, the mitigation of unintended effects in large-scale webgraphs can be achieved but comes with a trade-off on intervention performance.


\section{Discussion}

We find that link multiplicity weighting and link scheme removal interventions drive down the predicted traffic and domain ranking of unreliable news domains. While the design principles of generality and fairness mean that reliable domains are inevitably affected by these interventions, we find that the negative effects are mitigated---that is to say, low-quality sources disproportionately link to misinformation domains over reliable domains (\autoref{fig:link_scheme_target_dist}). 

As a result, link scheme removal is our most performant intervention. While we violate the principle of generality by identifying link schemes as sites that link heavily to unreliable domains, we use only a small sample of these labels (20\%). This method makes no assumptions about how link schemes link to reliable domains, yet we find that the distributions of their target domains are heavily skewed towards unreliable news sites (\autoref{fig:link_scheme_target_dist}).  Intervention results suggest that extending search algorithms such as PageRank to be link-scheme resilient is effective, and doing so will increase the cost of classic black-hat SEO strategies. This is needed, as black-hat SEO strategies are used disproportionately to promote misinformation \citep{misinformation_detection}. 

Instead of using reliability labels to identify low-quality link sources, the link multiplicity intervention relies on how webpages embed links. Source domains can embed many variants of the same URL in sidebars and information panes that persist across the site, and the resulting linkage patterns differentiate legitimate backlinking sites from those that implement black-hat SEO techniques like link-farming. As such, the link multiplicity intervention strikes an appropriate balance between generality, scalability, and performance, and we hope it can serve as a blueprint for future work on misinformation-resilient search engines. 

The effects of the small-scale interventions are estimates based on the predictions of our regression model. To explore the knock-on effects and demonstrate the robustness of these interventions, we evaluate the link scheme removal intervention in the context of a large-scale webgraph where PageRank can be computed directly. While modern search engines no longer rely on PageRank for their search rankings, we find that Ahrefs and SimilarWeb traffic estimates still correlate strongly with PageRank (\autoref{fig:simweb_comparison}). Furthermore, computing PageRank allows us to overcome the core limitation our small-scale interventions face---large-scale interventions do not employ predictive models, and as such are as close to a real-world experiment as if feasible within the scope of this study. That interventions performant on small-scale webgraphs (\autoref{tab:intervention_traffic_regressions}) can be translated to the large-scale setting (\autoref{tab:cc_webgraph_intervention_results}) demonstrates the effectiveness of our methodology for search-ranking intervention design and acts as a blueprint for future work in the area.

In exploring the unintended effects of our interventions in the large-scale webgraph setting, we discover additional categories of unreliable domains. We utilize this finding to develop the notion of multi-category link schemes as a subset of link schemes whose outlinks target various categories of unreliable domains. We visualize the log-pagerank distributions of these multi-category link schemes in (\autoref{fig:multi-domain-identification}). This method for identifying link schemes turns out to be quite precise, owing to the apparent abnormality of such a link structure, although we note that this precision comes at the cost of recall (and intervention effectiveness by extension). 

Anecdotally, we find that multi-category link schemes are a paid service, which somewhat explains the behavior of linking to multiple unrelated domain categories such as scam casinos and unreliable news sites. This distinction is crucial because unreliable news sites acquire farmed links through two primary methods: either the site owners procure these links through paid SEO services, or, unbeknownst to the site owner, the links are part of an information operation \citep{williams2023search}. As the quality of link scheme identification affects intervention effectiveness, we believe this distinction is an important avenue for future work.


\section{Limitations and Future Work}
Our methodology has three key limitations. First, we use \% change in traffic and rank as our evaluation metrics, when an argument can be made that the absolute change in these metrics is more important. However, the pre-post intervention distributions show a large discrepancy in the absolute traffic levels across label groups, with reliable news domains having much higher PageRank than unreliable domains (\autoref{fig:cc_webgraph_intervention_dist}). As a result, it would be extremely difficult to match absolute reductions in reliable news with absolute reductions in unreliable news, and interventions would not appear successful according to an absolute metric.

Second, our webgraph analysis uses static webgraphs. Assuming link-farming methods result in the bursty acquisition of links, link scheme identification would likely benefit from change-point detection methods that look for anomalies in links posted over time. Future research on link scheme identification with temporal webgraphs, where sites gain and lose links over time, is enabled by CommonCrawl's 14-year repository of webgraphs. A longitudinal study of traffic estimates, along the lines of work done for SEO spam detection in product reviews is also warranted \citep{bevendorff_is_nodate}.

Arguably the biggest limitation of this line of work stems from the opaqueness of search engines. As these algorithms are proprietary, we cannot use them to test our interventions, and must use approximations like PageRank or traffic estimates. Recent large-scale collaborations between Meta and misinformation researchers may provide a useful blueprint for how search engine companies could engage with researchers, ad tech companies, and other misinformation stakeholders to better understand the true impacts of interventions that aim to downrank unreliable content \cite{metanature2023,fbideology2023}. However, collaborations between academic institutions and search engine providers have not reached this level to date. Real-world evaluation of research into misinformation-resilient search rankings requires buy-in from search engine providers. 

Worse still, not all search engines prioritize countering misinformation in the same way. While large platforms like Google have greatly improved the reliability of their search results since 2016, other search engines still struggle in misinformation audit studies \cite{urman2022earth}. Confirmation bias is at work when those who are inclined to believe misinformation actively seek out alternative search engines. Recent work into the spread of `search directives' on social media uncovers how social media users are directed towards these platforms in search of conspiratorial content \citep{robertson2023identifying}. Unfortunately, algorithmic interventions aimed at downranking misinformation can only work insofar as those who consume such information actively use a system that employs that intervention. Future work into the factors that impact a user's preferred search engine is warranted. 

The perception of mainstream search engines as unfair and biased against certain political ideologies may be a contributing factor to these preferences. To this end, we propose that fairness-aware variants of the PageRank algorithm \citep{tsioutsiouliklis_fairness-aware_2021,pitoura_fairness_2022} can be extended to include political alignment as a protected attribute. A clear avenue for future research is to combine notions of fairness and safety in the design of search engine ranking interventions, motivated by the results detailed in \ref{fig:debias_results} which reveal fairness does not come at the cost of intervention quality. Despite the lack of political bias labels for many unreliable domains \autoref{fig:bias_by_reliability}), our debiasing results offer a promising direction.

\section{Conclusion}
This paper addresses the proliferation of unreliable news domains on search engines and presents a novel approach to penalize unreliable news domains while protecting reliable ones. We demonstrate that our interventions successfully reduce the search rankings and traffic received by unreliable news domains while maintaining relatively high traffic and rankings for reliable news domains. We develop a principled approach to the design of such interventions, considering fairness, generality, cost (for an adversary), and scalability. Following this, we propose two interventions, link scheme removal, and link multiplicity weighting, and we evaluate them on small-scale webgraph data using a regression model that predicts changes in traffic based on simulated modifications to the backlink networks of labeled news domains. We then generalize the link scheme removal interventions to run experiments on large-scale webgraph data consisting of 93.9 million domains and calculate PageRank differences following our interventions.

The effectiveness of the interventions is demonstrated through the reduction of traffic to misinformation sites compared to reliable sites. The Link Scheme Removal intervention balances a 35\% drop in unreliable traffic with an 11\% drop in reliable traffic. By combining this with the multiplicity intervention, we achieve a drop in estimated unreliable traffic of 32\%, with only a 5\% drop in traffic to reliable news domains. These results demonstrate a favorable trade-off between reducing the influence of misinformation domains and minimizing collateral damage to reliable domains. This tradeoff remains favorable when the interventions are evaluated in the large-scale webgraph setting, with the best performing `fair' intervention achieving a 10\% decrease in unreliable PageRank centrality, at the cost of a 2\% decrease for reliable news domains.

To mitigate concerns over their impact on reliable domain traffic, we demonstrate that these interventions can be tuned to prevent estimated traffic loss to reliable sites. The interventions are flexible enough that tuning them can change the effect from downranking unreliable sites to promoting reliable ones. For the large-scale webgraph setting, we propose a link-scheme identification method that reduces the unintended effects of running edge removal interventions.
Finally, by debiasing the dataset and showing that intervention performance is unchanged, we provide evidence for the possibility of a safe \textit{and} fair PageRank that actively prevents the dissemination of misinformation in an unbiased manner despite the strong correlation between political alignment and reliability. Along with the principle of generality, we believe that such algorithmic fairness is a necessary condition for public trust and acceptance. 

Our main contribution is the development of targeted strategies to enhance the trustworthiness and quality of search rankings. The proposed interventions have the potential to benefit users and the broader digital community by combating the spread of misinformation and fostering a more reliable online information ecosystem. These findings serve as a foundation for improving the resilience of ranking-based information retrieval systems.

\section{Acknowledgements}
This work was supported in part by the Office of Naval Research grant (N000141812106) and the Knight Foundation. Additional support was provided by the Center for Computational Analysis of Social and Organizational Systems (CASOS) at Carnegie Mellon University. The views and conclusions contained in this document are those of the authors and should not be interpreted as representing the official policies, either expressed or implied, of the Knight Foundation, Office of Naval Research, or the U.S. Government.

\bibliographystyle{ACM-Reference-Format}
\bibliography{sample-base, PR, Casinos}

\begin{appendices}
\appendix
\section{Neural Baselines}\label{app:NB}
Our vanilla neural network model is simply one affine layer with a dropout of 0.5 and a ReLU activation function, followed by a linear projection to a 1-dimensional subspace. The GNN models both contain a single Graph Convolutional Layer \cite{kipf2016semi} which can be formally written as: 

\begin{equation}
H = \sigma(\tilde{D}^{-1/2}\tilde{A}\tilde{D}^{-1/2}XW)
\end{equation}

where $\sigma$ is a ReLU activation, $\tilde{A}$ is the sum of the Adjacency Matrix $A$ and its identity matrix $I$, $\tilde{D}$ is the degree matrix of $\tilde{A}$, $X$ are features of nodes and their backlinking domains, and $W$ is a learnable weight matrix. In the GNNs, $A$ is a matrix containing all reliability-labeled domains and edges from their 10 most frequent backlinking domains. We consider an unweighted GNN ($GNN_{uw}$) where elements of $A$ are binary and a weighted GNN ($GNN_{w}$) where elements of $A$ are not constrained to be binary. Given the large variance in backlink volumes, we log-transform non-zero weights in the weighted setting. All models are trained with an Adam optimizer with an initial learning rate of 1e-2, Cross Entropy Loss, a Cosine Annealing Scheduler, and Early Stopping with a patience of 100. 

\begin{figure}[!h]
\centering
\includegraphics[width=.58\linewidth]{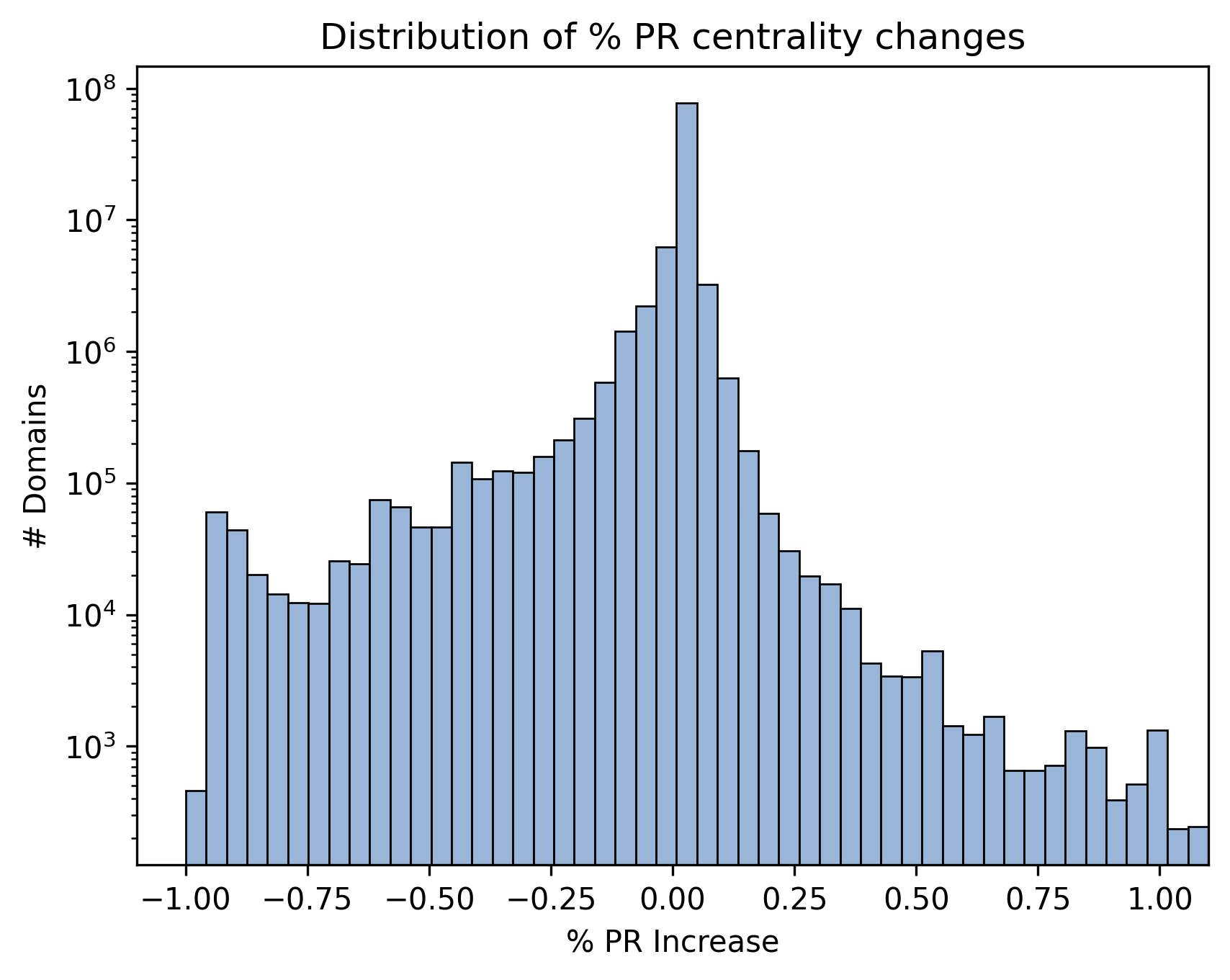}
\caption{Logged distribution of PR changes across all domains for the ATR extended edge removal intervention shows limited unintended effects. A long-tail of domains whose PR increases by more than 100\% is excluded.}
\label{fig:pr_reduction_dist}
\end{figure}

\end{appendices}

\end{document}